\documentclass{article}
\usepackage{cprotect}
\usepackage{multirow}
\usepackage{booktabs}
\usepackage{algorithm}
\usepackage{algpseudocode}
\usepackage{amsmath}
\usepackage{amssymb}
\usepackage{graphicx}
\usepackage{subcaption}
\usepackage[margin=1in]{geometry}
\usepackage[nolist,nohyperlinks]{acronym}
\usepackage{url} 
\usepackage{tikz}
\usetikzlibrary{arrows.meta, shapes.geometric, positioning, fit}
\usepackage{authblk}
\usepackage{enumitem}

\title{FAMST: Fast Approximate  Minimum Spanning Tree Construction for Large-Scale and High-Dimensional Data}

\author[1]{Mahmood K. M. Almansoori}
\author[1,2]{Miklos Telek}

\affil[1]{\small \textit{Dept.\ of Networked Systems and Services, Budapest University of Technology and Economics, Budapest, Hungary}}
\affil[2]{\small \textit{HUNREN–BME Information Systems Research Group, Budapest University of Technology and Economics, Budapest, Hungary}}

\affil[ ]{\texttt{\{almansoori,telek\}@hit.bme.hu}}

\date{}

\begin{document}

\maketitle

\begin{abstract}
We present \ac{FAMST}, a novel algorithm that addresses the computational challenges of constructing \acp{MST} for large-scale and high-dimensional datasets. \ac{FAMST} utilizes a three-phase approach: \ac{ANN} graph construction, \ac{ANN} inter-component connection, and iterative edge refinement. For a dataset of $n$ points in a $d$-dimensional space, \ac{FAMST} achieves $\mathcal{O}(dn \log n)$ time complexity and $\mathcal{O}(dn + kn)$ space complexity when $k$ nearest neighbors are considered, which is a significant improvement over the $\mathcal{O}(n^2)$  time and space complexity of traditional methods.

Experiments across diverse datasets demonstrate that \ac{FAMST} achieves remarkably low approximation errors while providing speedups of up to 1000$\times$ compared to exact \ac{MST} algorithms. We analyze how the key hyperparameters, $k$ (neighborhood size) and $\lambda$ (inter-component edges), affect performance, providing practical guidelines for hyperparameter selection. \ac{FAMST} enables \ac{MST}-based analysis on datasets with millions of points and thousands of dimensions, extending the applicability of \ac{MST} techniques to problem scales previously considered infeasible.
\end{abstract}

\section{Introduction}
\label{sec:intro}
\acp{MST} are fundamental structures in data analysis and machine learning, offering efficient ways to capture global relationships among data points. They are central to tasks such as hierarchical clustering, outlier detection, dimensionality reduction, and manifold learning~\cite{xu2015comprehensive, mohapatra2022survey}. The \ac{MST} of a dataset ensures all points are connected with the minimum possible total edge weight, making it a powerful tool for revealing intrinsic data structure.

While classical \ac{MST} algorithms, such as Kruskal’s and Prim’s, are computationally efficient for extracting a spanning tree from a given graph, the true computational bottleneck lies in the construction of the complete graph itself. 
Traditional \ac{MST} workflows typically begin by forming a complete weighted graph, where each edge represents the pairwise distance between data points.
This step alone involves computing all $\mathcal{O}(n^2)$  pairwise distances, leading to quadratic time and space complexity. As datasets grow to millions of points or become high-dimensional, this exhaustive distance computation becomes prohibitively expensive.

Moreover, common spatial acceleration structures (e.g., KD-trees or Ball trees), which can mitigate the cost of distance calculations in low-dimensional settings, degrade in performance as dimensionality increases due to the curse of dimensionality~\cite{march2010fast}. As a result, the main challenge is not in applying \ac{MST} algorithms like Kruskal's, but in avoiding the need to compute and store the full graph in the first place.


To address this challenge, recent work has explored approximate strategies that reduce the number of distance evaluations by operating on clustering-based methods or sparse neighborhood graphs such as \ac{$k$NN} or \ac{ANN} graphs. While effective in lowering computational costs, these methods often produce disconnected components, and the key challenge becomes how and when to reconnect them. There are two primary approaches: one can first construct an approximate \ac{MST} forest over each component and then attempt to augment this forest to form an approximate global \ac{MST} \cite{veldt2025approximate}, or—alternatively—connect the \ac{ANN} graph into a single component first, then extract the approximate \ac{MST} directly. The latter strategy forms the core of our proposed approach.

In this paper, we present \ac{FAMST}, a novel algorithm designed to efficiently construct high-quality \ac{MST} approximations for large-scale and high-dimensional datasets. \ac{FAMST} employs a three-phase pipeline: (1) construction of a sparse \ac{ANN} graph, (2) strategic connection of disconnected components, and (3) iterative refinement of inter-component edges based on local neighborhood exploration. Finally, Kruskal’s algorithm is applied to extract an \ac{MST} that closely approximates the optimal one.

Our key contributions are as follows:
\begin{itemize}
  \item We propose a scalable algorithm for approximate \ac{MST} construction with near-linear time complexity $\mathcal{O}(dn \log n)$ and memory usage $\mathcal{O}(dn + kn)$, significantly improving over traditional $\mathcal{O}(n^2)$ methods.
  \item We introduce a lightweight yet effective edge refinement strategy that improves approximate \ac{MST} quality by reducing unnecessary long-range edges.
  \item We conduct extensive experiments across real and synthetic datasets, demonstrating up to 1000$\times$ speedup over exact \ac{MST} algorithms with minimal approximation error.
  \item We provide a detailed analysis of the algorithm’s hyperparameters and scalability, offering practical guidelines for deployment on modern data.
\end{itemize}

\ac{FAMST} addresses a critical gap in the computation of \acp{MST}, offering a practical solution for scenarios where the scale of data makes exact methods computationally infeasible. Our approach balances approximation quality and computational efficiency, enabling \ac{MST}-based analyses on modern large-scale and high-dimensional datasets.

The remainder of this paper is organized as follows. Section \ref{sec:intro} reviews related work in both exact and approximate \ac{MST} algorithms. Section \ref{sec:methodology} presents the \ac{FAMST} algorithm, detailing its three-phase pipeline for efficient \ac{MST} approximation. In Section \ref{sec:complexityAnalysis}, we analyze the theoretical computational and space complexity of \ac{FAMST}. Section \ref{sec:experiments} offers an extensive empirical evaluation across diverse datasets, comparing \ac{FAMST} with existing approaches in terms of speed and accuracy. Section \ref{sec:hyperparameters} explores the influence of key hyperparameters on performance and provides practical guidelines. Section \ref{sec:furtherImprovements} discusses potential enhancements and future work. Finally, Section \ref{sec:conclusion} concludes the paper by summarizing our contributions and key findings.

\section{Related work}
\label{sec:related}
Classical \ac{MST} algorithms suffer from poor scalability, especially in high-dimensional or large-scale settings, where computing all pairwise distances becomes computationally infeasible. This section reviews prior work in two main categories: exact \ac{MST} algorithms and approximate \ac{MST} methods.

\subsection{Exact \ac{MST}}
Constructing an \ac{MST} from a dataset of $n$ points in a $d$-dimensional space involves two main phases. The first step is to build a fully connected, weighted undirected graph in which each vertex corresponds to a data point, and the edge weights reflect pairwise similarities (or distances) between points. The second step involves applying an  \ac{MST} algorithm to extract the tree that spans all vertices with the minimal total edge weight.

Among the most widely used algorithms for this task are Kruskal’s, Prim’s, and Borůvka’s algorithms  \cite{mohapatra2022survey}. 
Kruskal’s algorithm, in particular, follows a greedy strategy: it begins by initializing each node in its own disjoint set, then sorts all graph edges in ascending order of weight. Edges are added one by one to the  \ac{MST} if they do not introduce a cycle, and the sets of their endpoints are merged. This process continues until the tree contains exactly $n-1$  edges, where $n$ is the number of nodes.

While these algorithms are efficient in processing the graph structure itself, the primary computational bottleneck lies in constructing the complete graph, which requires computing $\mathcal{O}(n^2)$ pairwise distances. This makes exact \ac{MST} construction intractable for large-scale or high-dimensional datasets.

The Dual-Tree Boruvka algorithm was introduced in \cite{march2010fast}  for computing the \ac{EMST} of a set of points in Euclidean space.
This algorithm enhances the classical Boruvka method by integrating dual-tree traversal, which simultaneously navigates two spatial trees (e.g., KD-trees) to eliminate redundant distance calculations. 
In low-dimensional settings, it achieves time complexity of $\mathcal{O}(n~log~n)$, significantly outperforming the naive $\mathcal{O}(n^2)$ approach that computes all pairwise distances.
While \ac{EMST} is highly effective in low to moderate dimensions,  the \ac{EMST} is not effective for high-dimensional data due to the curse of dimensionality. For instance, when KD-trees or Ball trees are applied to high-dimensional data, most of the points in the tree must be evaluated, resulting in an efficiency that is no better than brute force search \cite{liu2006new, ram2012nearest, ram2019revisiting, gallego2020insights}. 

These limitations make exact \ac{MST} computation impractical for many real-world scenarios, motivating an extensive research effort on approximate \ac{MST} methods designed to trade off a small loss in accuracy for massive improvements in scalability.

\subsection{Approximate \ac{MST}}

Given the computational challenges of exact \ac{MST} construction, a variety of approximate \ac{MST} algorithms have emerged, many of which rely on clustering or sparse neighborhood graphs to reduce complexity. These approaches replace the fully connected graph with a clusters-based, \ac{$k$NN}  or \ac{ANN} graph, dramatically reducing the number of required distance computations.

A common approach computes \acp{MST} on data partitions and then extends them to the full set \cite{wang2009divide, zhong2015fast, jothi2018fast, ghosh2024kmist}. Both computational cost and accuracy depends on the quality of the partitioning mechanism—whether clustering algorithms or spatial indexes such as k-d trees—and these mechanisms often lose efficiency and precision as dimensionality grows.  


Fast Parallel Algorithms for Euclidean \ac{MST} and HDBSCAN$^{*}$ were introduced in \cite{parallel2021fast} to accelerate both tasks by using a well-separated pair decomposition (WSPD) method \cite{callahan1995decomposition}. However, a WSPD contains $\mathcal{O}(s^{d}n$) pairs ($s$ denote the separation factor of WSPD;  $s > 0$), so its constant grows exponentially with the dimension $d$ \cite{callahan1995decomposition, govindarajan2006efficient}. 
Consequently,the Euclidean \ac{MST} proposed in \cite{parallel2021fast} is highly effective in low- and moderate-dimensional Euclidean spaces, but the computational time and memory blow-up make it unattractive once $d$ rises past a few dozen \cite{almansoori2025fast}.




The connection between nearest neighbor graphs and \acp{MST} has inspired several efficient approximation techniques. A \ac{$k$NN} graph, which connects each point to its $k$ closest neighbors, often captures many of the edges present in the exact \ac{MST}. This observation has led to algorithms that use \ac{$k$NN} graphs as starting points for \ac{MST} approximation.

Graph-based approaches for \ac{ANN} search, such as \ac{NSW} graphs \cite{malkov2014approximate} and  \ac{HNSW} graphs \cite{malkov2018efficient}, have demonstrated impressive performance for similarity search in high-dimensional spaces. These methods construct navigable graph structures that enable efficient nearest neighbor queries with logarithmic complexity. However,  the \ac{HNSW} approach demands considerable memory resources \cite{malkov2018efficient}.

FISHDBC method \cite{dell2019fishdbc} utilized \ac{HNSW} to construct an approximate \ac{MST} for scalable, density-based clustering. FISHDBC builds an \ac{MST} incrementally using candidate edges derived from distance evaluations while constructing the \ac{HNSW} index. As new data points are added, distances between neighboring points are computed and collected as edge candidates. Periodically, the algorithm merges these candidate edges into the current \ac{MST} using Kruskal’s algorithm, maintaining a spanning forest that may temporarily consist of disconnected components. The authors argue—based on theoretical justification—that missing edges can be assumed to have infinite weight and thus do not affect the final hierarchical clustering. This method prioritizes building a structure suitable for clustering rather than computing the best possible approximation of the \ac{MST} itself.

NN-Descent \cite{dong2011efficient} is a widely used algorithm for approximate \ac{$k$NN}  graph construction, known for its simplicity, efficiency, and applicability to various distance metrics. It iteratively refines a randomly initialized \ac{$k$NN}  graph by leveraging the observation that “a neighbor of a neighbor is likely to be a neighbor,”  which allows it to focus comparisons on promising candidates rather than exhaustively evaluating all possible pairs.

The PyNNDescent library \cite{pynndescent}  was introduced as an optimized and more practical implementation of NN-Descent. PyNNDescent improves both runtime performance and usability by incorporating efficient data structures, parallelism, and better memory handling. It also supports a wide range of distance metrics and integrates easily with Python-based machine-learning workflows.

GNND (GPU-based NN-Descent) \cite{wang2021fast} is a GPU-accelerated variant of  NN-Descent. GNND preserves the core ideas of NN-Descent but introduces several GPU-oriented optimizations, including a selective update mechanism and memory-efficient sampling strategies, which reduce data movement between GPU cores and global memory. Additionally, they propose a GPU-based graph merging technique that enables efficient construction of \ac{$k$NN} graphs for datasets exceeding GPU memory capacity.

Several researchers have proposed the use of \ac{ANN} graphs specifically for \ac{MST} approximation \cite{dell2019fishdbc, Naidoo2019, veldt2025approximate}. However, a key challenge with these approaches is ensuring connectivity, as \ac{$k$NN} graphs often consist of multiple disconnected components, particularly for reasonably small values of $k$.


MiSTree \cite{Naidoo2019} is a graph-based approach that constructs an approximate \ac{MST} by first generating a \ac{$k$NN} graph and then applying Kruskal’s algorithm to extract the spanning structure. MiSTree relies on exact nearest neighbor search, which limits its scalability to low or moderately sized and low-dimensional datasets. Moreover and importantly, it does not guarantee a fully connected \ac{MST}: if the initial \ac{$k$NN} graph is disconnected due to an insufficient neighborhood size $k$, the resulting output is a forest of \acp{MST} rather than a single tree.

Another \ac{MST} approximation method was proposed by \cite{veldt2025approximate} for arbitrary metric spaces, introducing the \ac{MFC} problem. 
Starting from an initial \ac{MST} forest—a set of disjoint \acp{MST} obtained cheaply (e.g., from a kNN graph or clustering then applying exact \ac{MST} on each cluster)—their MFC algorithm randomly selects one representative per component then queries distances between every data point and the representatives to construct inter-clusters (inter-components) edges. Finally, Kruskal's algorithm is applied to obtain the final approximate MST. 

Our work builds upon these foundations by introducing a systematic approach to connect \ac{ANN} graph components and iteratively refine these connections to minimize the total weight of the approximate spanning tree. Our refinement process strategically improves edge selection based on local neighborhood exploration, leading to high-quality \ac{MST} approximations even for challenging high-dimensional datasets.

\section{Methodology}
\label{sec:methodology}

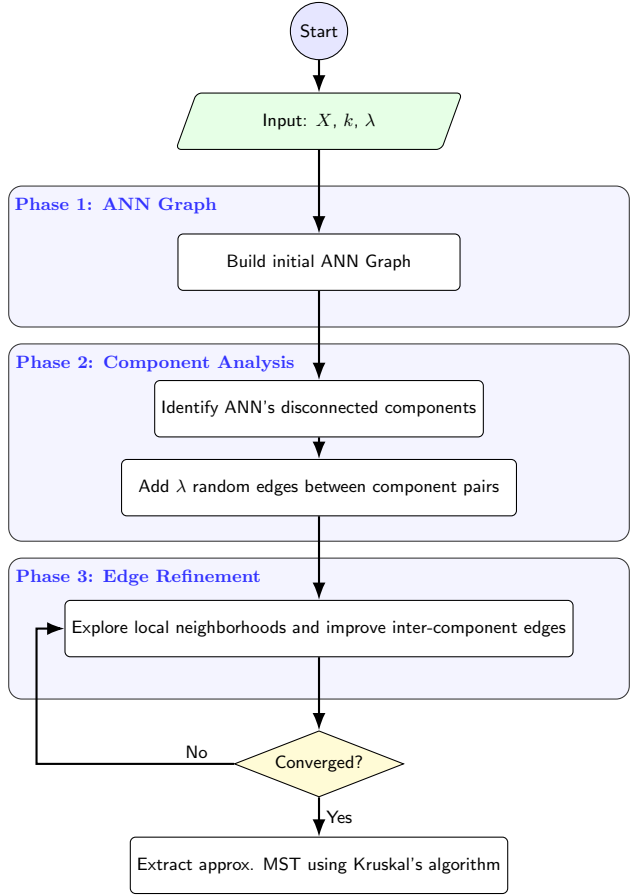
\begin{figure}
    \centering 
    
    \begin{tikzpicture}[
        font=\sffamily\small,
        box/.style={draw, fill=white, rounded corners=2pt, minimum height=1cm, align=center},
        io/.style={box, fill=green!10, trapezium, trapezium left angle=70, trapezium right angle=110},
        decision/.style={diamond, draw, fill=yellow!20, aspect=2, inner sep=1pt, minimum width=3cm},
        phase/.style={draw, fill=blue!5, rounded corners=5pt, opacity=0.8},
        arrow/.style={-Latex, thick},
        phaseLabel/.style={font=\bfseries\small\color{blue!80}}, scale=0.75, every node/.style={transform shape},
    ]
    
    \node[phase, minimum width=11cm, minimum height=2.5cm] (phase1) at (0,3.3) {};
    \node[phaseLabel] at (-3.6,4.2) {Phase 1: \ac{ANN} Graph};

    \node[phase, minimum width=11cm, minimum height=3.5cm] (phase2) at (0,0) {};
    \node[phaseLabel] at (-2.9,1.4) {Phase 2: Component Analysis};
    
    \node[phase, minimum width=11cm, minimum height=2.5cm] (phase3) at (0,-3.3) {};
    \node[phaseLabel] at (-3.2,-2.4) {Phase 3: Edge Refinement};

    \node[circle, draw, fill=blue!10, minimum size=1cm] (start) at (0,7.3) {Start};
    \node[io, minimum width=4cm] (input) at (0,5.7) {Input: $X$, $k$, $\lambda$};
    
    \node[box, minimum width=5cm] (ann) at (0,3.2) {Build initial \ac{ANN} Graph};
    
    \node[box, minimum width=5cm] (components) at (0,0.6) {Identify ANN's disconnected components};
    \node[box, minimum width=7cm] (lambdaedges) at (0,-0.8) {Add $\lambda$ random edges between component pairs};
    
    \node[box, minimum width=7cm] (refine) at (0,-3.3) {Explore local neighborhoods and improve inter-component edges};
    \node[decision] (converged) at (0,-5.7) {Converged?};
    \node[box, minimum width=6cm] (mst) at (0,-7.5) {Extract approx. \ac{MST} using Kruskal's algorithm};
    
    \draw[arrow] (start) -- (input);
    \draw[arrow] (input) -- (ann);
    \draw[arrow] (ann) -- (components);
    \draw[arrow] (components) -- (lambdaedges);
    \draw[arrow] (lambdaedges) -- (refine);
    \draw[arrow] (refine) -- (converged);
    \draw[arrow] (converged) -- node[right]{Yes} (mst);
    
    \draw[arrow] (converged.west) -- ++(-3.5,0) 
        node[above left=-1pt and -90pt]{No} -- ++(0,2.4) -- (refine.west);
    
    \end{tikzpicture}
    \caption{\ac{FAMST} pipeline.}
    \label{fig:FAMST_pipeline}
    
\end{figure}

The core idea of our approach is to: (1) build a sparse graph that captures the essential local structure of the data using \ac{ANN} search, (2) ensure connectivity through inter-component edge addition, and (3) iteratively refine these inter-component connections to improve the quality of the approximate one-component \ac{ANN} graph. Finally, (4)  \ac{MST} extraction using Kruskal's algorithm (or any other method) on the connected \ac{ANN} graph (see Figure \ref{fig:FAMST_pipeline}). This process yields an approximate \ac{MST} whose weight approaches that of the true \ac{MST} while avoiding the quadratic complexity of exact methods.

\begin{figure*}
     \centering
     \begin{subfigure}[b]{0.32\textwidth}
         \centering
         \includegraphics[width=\textwidth]{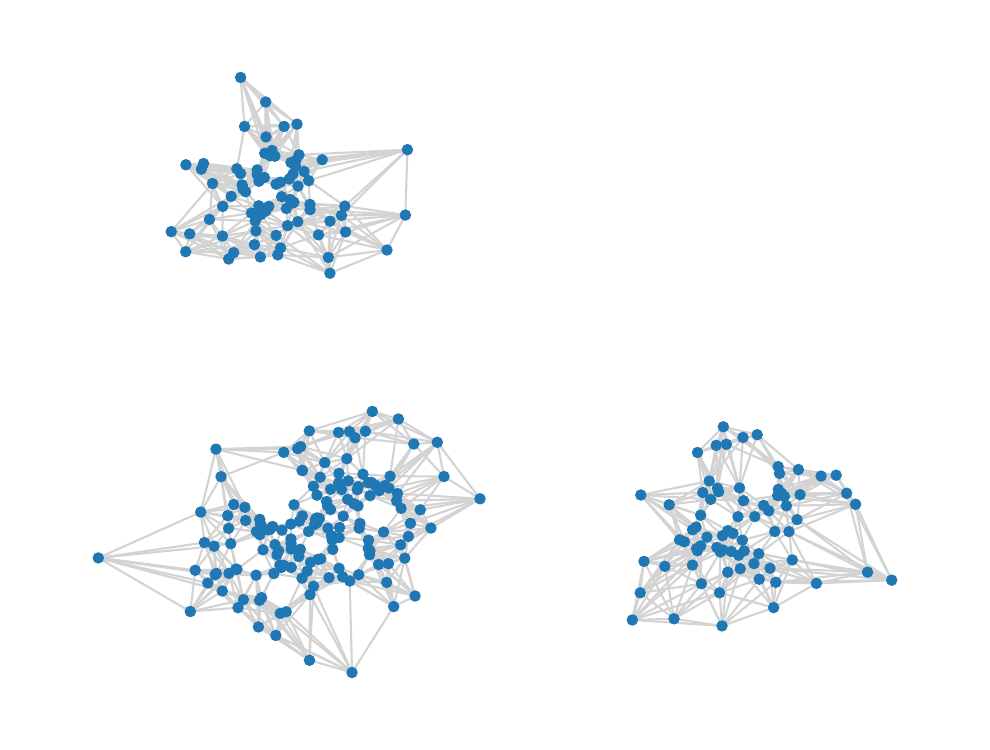}
         \caption{Disconnected ANN}
         \label{fig:disconnectedANN}
     \end{subfigure}
     \hfill
     \begin{subfigure}[b]{0.32\textwidth}
         \centering
         \includegraphics[width=\textwidth]{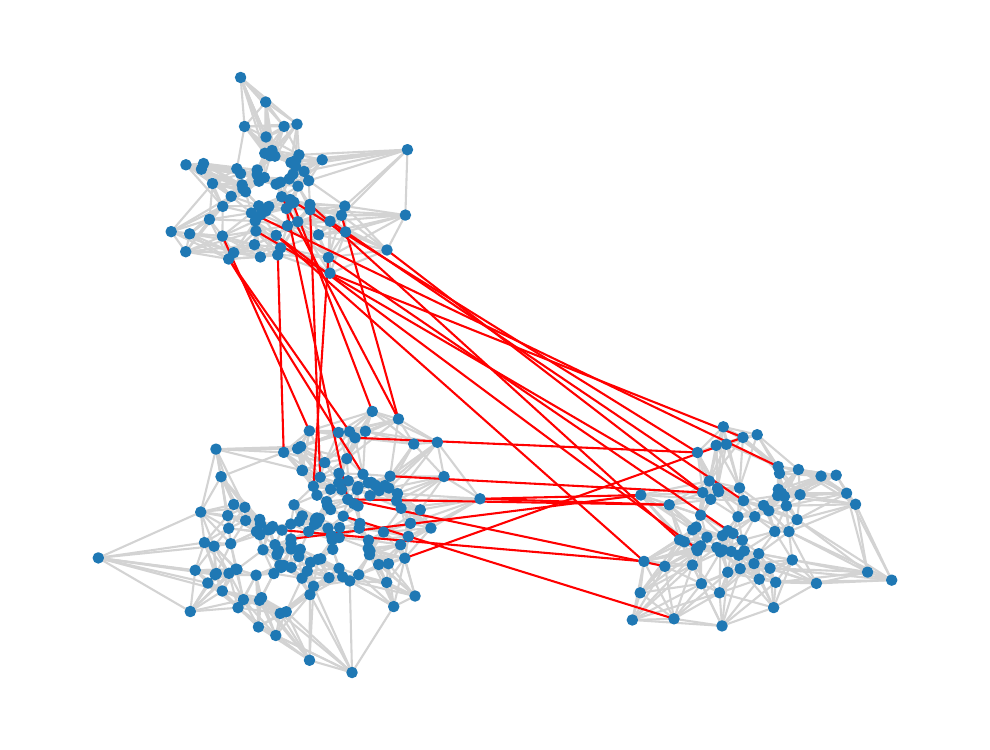}
         \caption{Connected ANN}
         \label{fig:connectedANN}
     \end{subfigure}
     \hfill
     \begin{subfigure}[b]{0.32\textwidth}
         \centering
         \includegraphics[width=\textwidth]{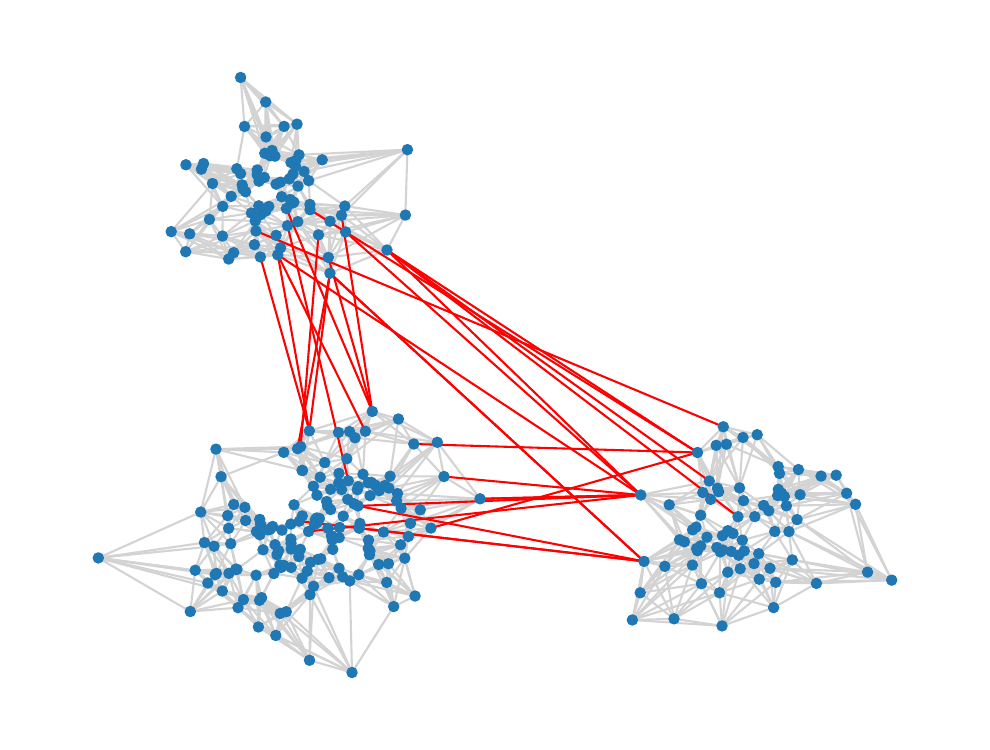}
         \caption{Refinement 01}
         \label{fig:refinement01}
     \end{subfigure}
     \hfill
     \begin{subfigure}[b]{0.32\textwidth}
         \centering
         \includegraphics[width=\textwidth]{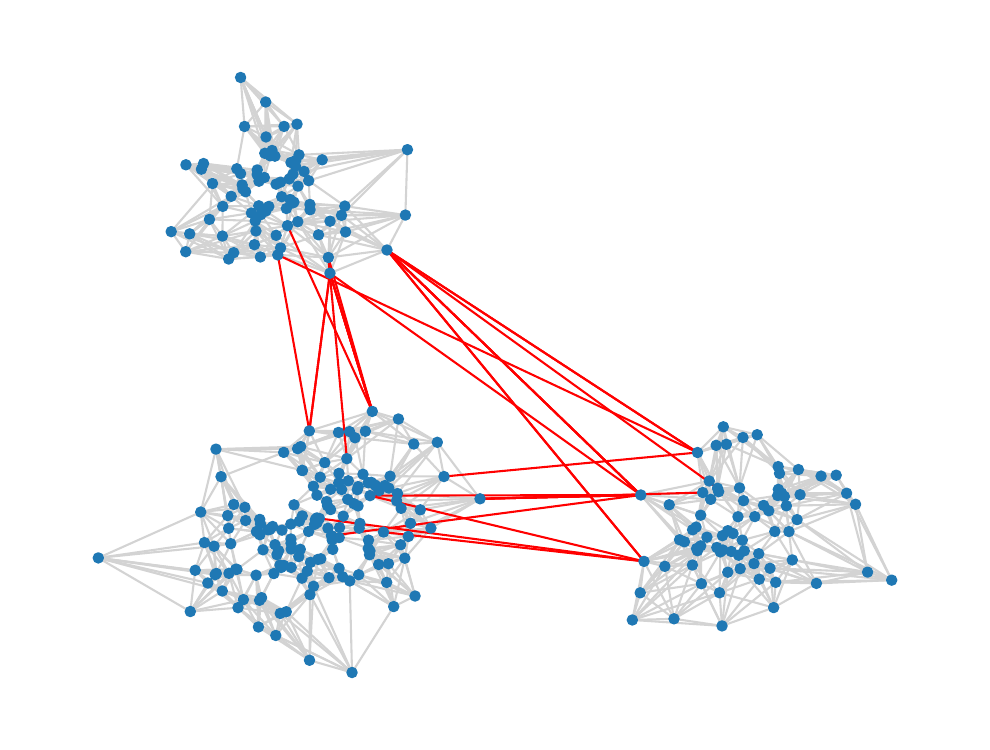}
         \caption{Refinement 02}
         \label{fig:refinement02}
     \end{subfigure}
     \hfill
     \begin{subfigure}[b]{0.32\textwidth}
         \centering
         \includegraphics[width=\textwidth]{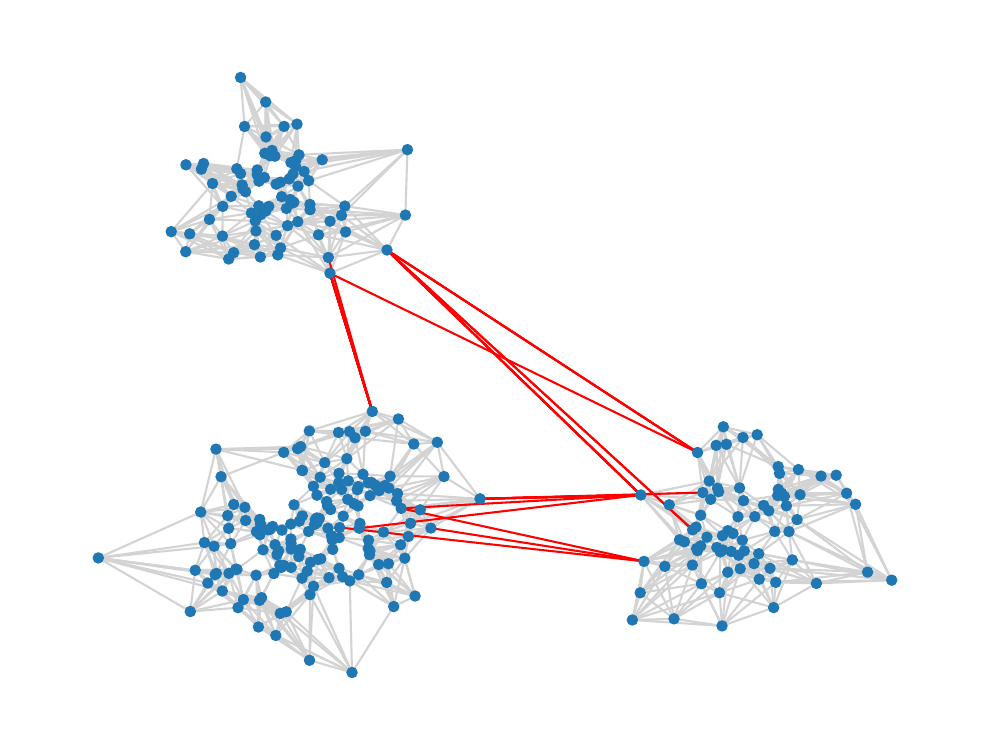}
         \caption{Refinement 03}
         \label{fig:refinement03}
     \end{subfigure}
     \hfill
     \begin{subfigure}[b]{0.32\textwidth}
         \centering
         \includegraphics[width=\textwidth]{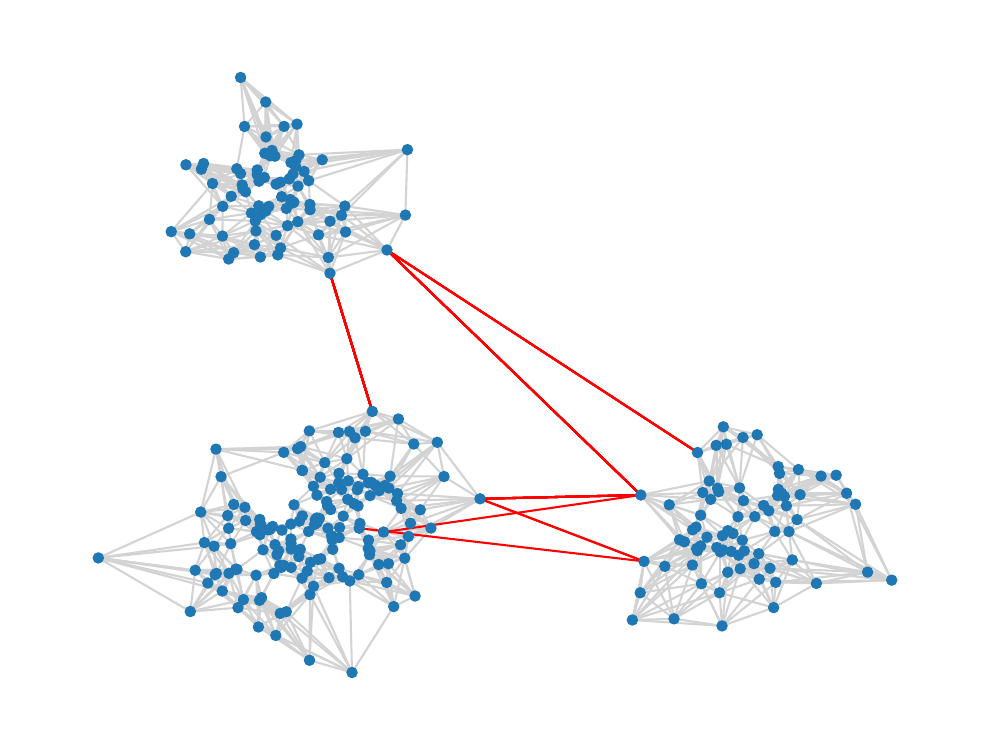}
         \caption{Refinement 04}
         \label{fig:refinement04}
     \end{subfigure}
     \begin{subfigure}[b]{0.32\textwidth}
         \centering
         \includegraphics[width=\textwidth]{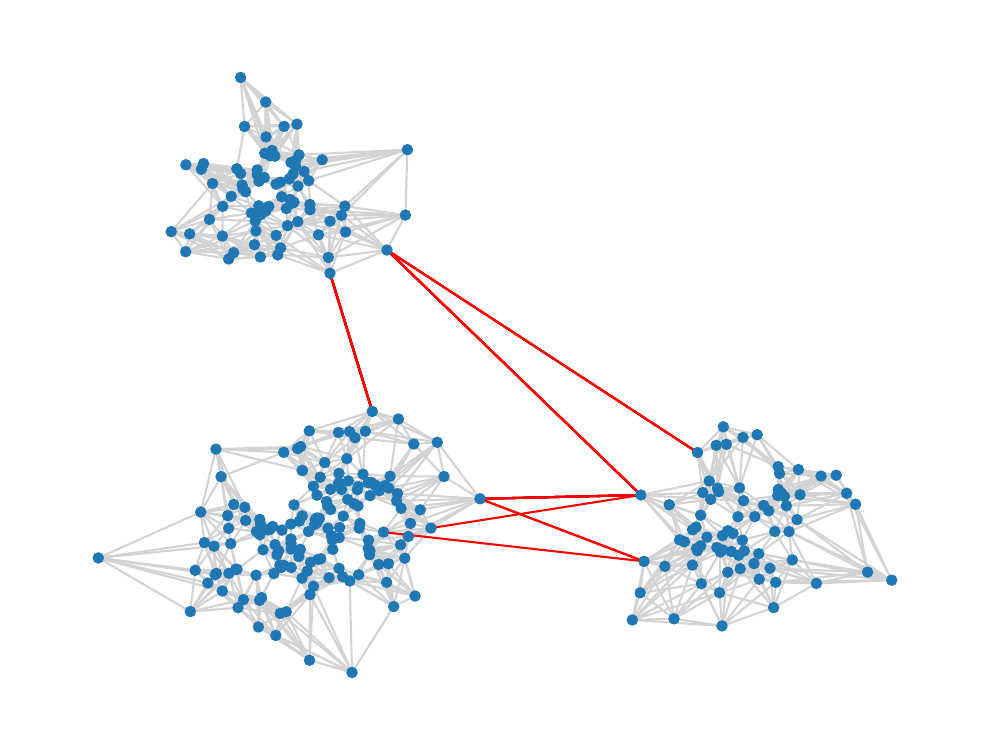}
         \caption{Refinement 05}
         \label{fig:refinement05}
     \end{subfigure}
     \begin{subfigure}[b]{0.32\textwidth}
         \centering
         \includegraphics[width=\textwidth]{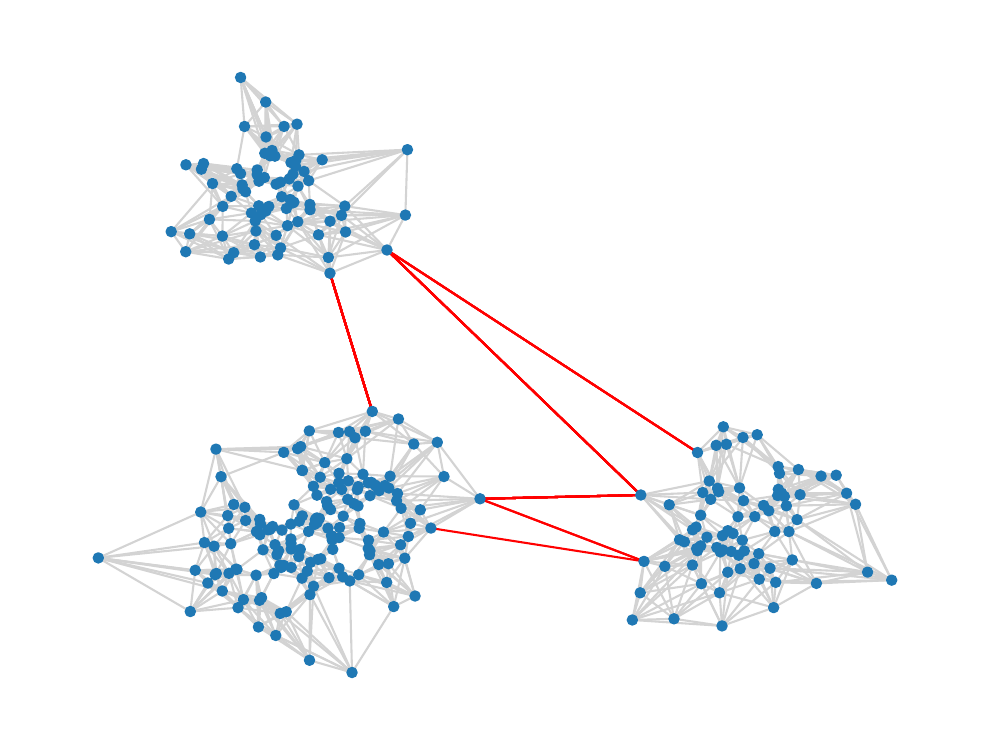}
         \caption{Refinement 06}
         \label{fig:refinement06}
     \end{subfigure}
     \begin{subfigure}[b]{0.32\textwidth}
         \centering
         \includegraphics[width=\textwidth]{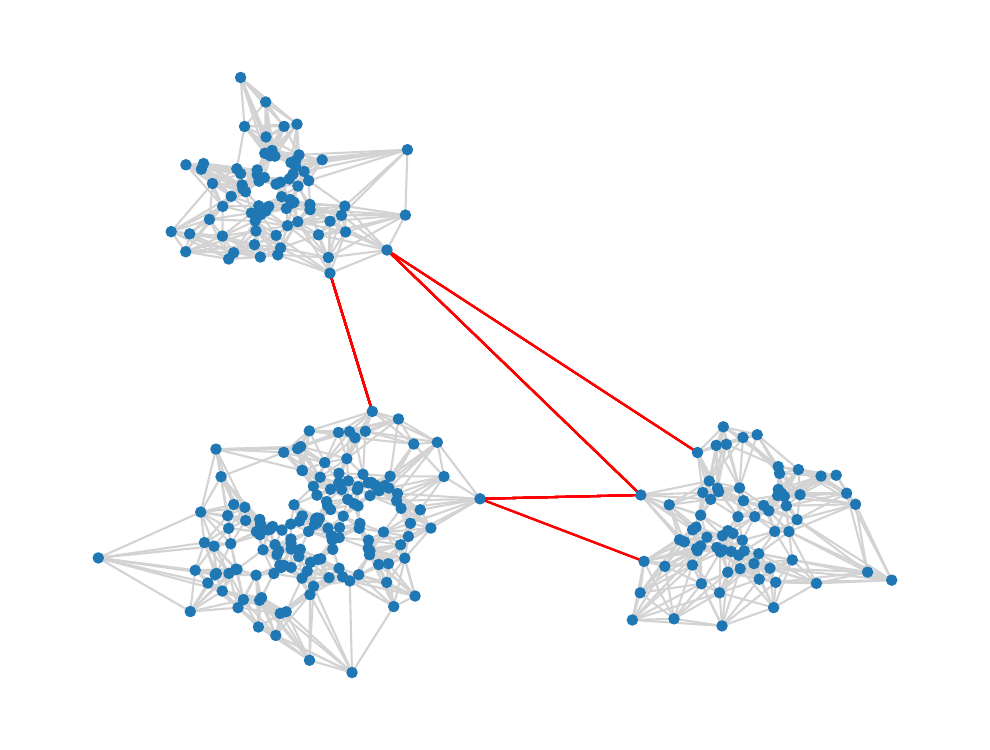}
         \caption{Refinement converge}
         \label{fig:refinementConverge}
     \end{subfigure}
     \begin{subfigure}[b]{0.32\textwidth}
         \centering
         \includegraphics[width=\textwidth]{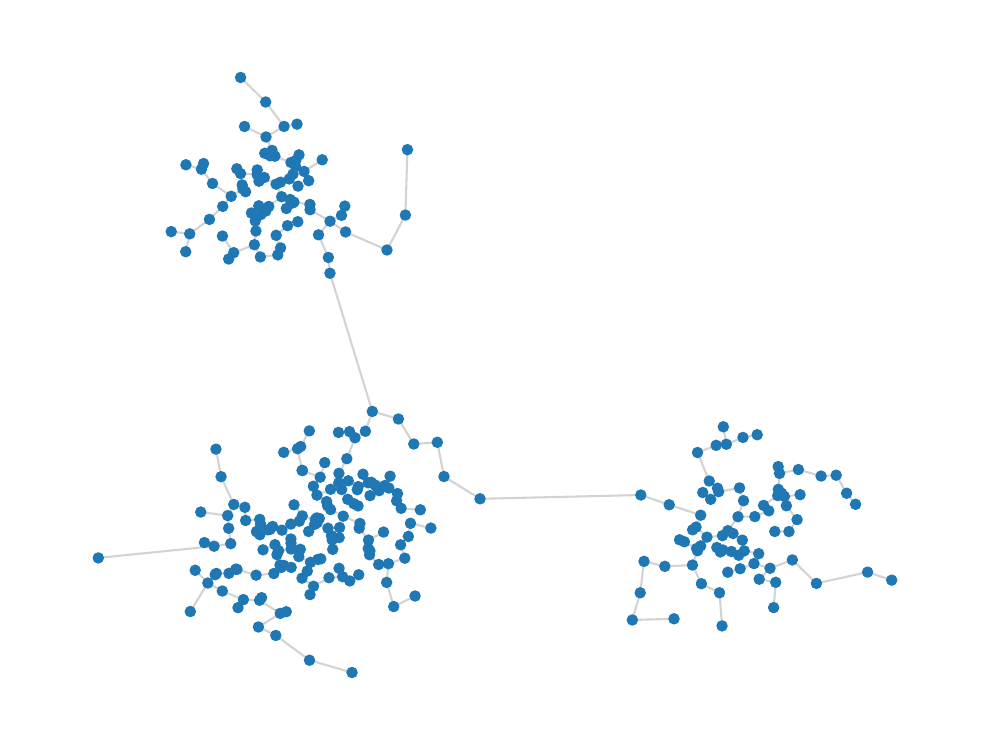}
         \caption{Approx. \ac{MST}}
         \label{fig:finalMST}
     \end{subfigure}
        \caption{Steps of obtaining approx. \ac{MST}. }
        \label{fig:toyExample}
\end{figure*}

The refinement process is illustrated in Figure \ref{fig:toyExample}, showing how the initial connections between components (subfigure \ref{fig:connectedANN}) are progressively improved through multiple refinement iterations until convergence to the final connected \ac{ANN} (subfigure \ref{fig:refinementConverge}).

The proposed method consists of three main phases:

\begin{itemize}[label={}, leftmargin=2cm, labelsep=0.5cm]
  \item[\textbf{Phase 1}] \textbf{\ac{ANN} graph construction}: Build an initial \ac{ANN} graph
  \item[\textbf{Phase 2}] \textbf{Component analysis and connection}: Identify disconnected components and establish preliminary inter-component connections
  \item[\textbf{Phase 3}] \textbf{Edge refinement}: Iteratively improve the quality of inter-component connections
\end{itemize}

Algorithm \ref{alg:fullprocess} outlines the overall process of \ac{FAMST} for constructing an approximate \ac{MST}. The initial step involves computing the $k$ approximate nearest neighbors for each data point using efficient \ac{ANN} search techniques. This produces a sparse graph with $\mathcal{O}(kn)$ edges, forming a directed weighted graph that may contain multiple disconnected components. This graph has significantly fewer edges than the $\mathcal{O}(n^2)$ edges in a complete graph.
For \acp{ANN} graph construction (phase 1), any algorithm could be used as long as it produces neighbor lists ($\mathcal{N}$) and corresponding distances ($ \mathcal{D}$). In this work, we utilized the PyNNDescent library to obtain the \acp{ANN} graph.

\begin{algorithm*}
\caption{\ac{FAMST}}
\label{alg:fullprocess}
\begin{algorithmic}[1]
\Require $X$, number of neighbors $k$, number of random edges \( \lambda \)
\Ensure Approximate \ac{MST} $\mathcal{T}$

\State \( \mathcal{N}, \mathcal{D} \gets \text{GetANNs}(X, k) \) \Comment{Neighbors and Distances each $n \times k$}
\State \( C, \mathcal{G} \gets \text{Components}(\mathcal{N}) \) \Comment{\acp{ANN} Components and Undirected graph (Algorithm \ref{alg:find_components})}
\State \( E, E_C \gets \text{RandomEdges}(X, C, \lambda) \) \Comment{Add inter-component edges (Algorithm \ref{alg:add_random_edges})}

\Repeat
    \State \( E, \Delta \gets \text{RefineEdges}(X, \mathcal{G}, C, E, E_C) \) \Comment{Inter-component edges refinements (Algorithm \ref{alg:refine_edges})}
\Until{\( |\Delta| = 0 \)} \Comment{Iterate until no further changes}

\State \( E_{\text{final}} \gets E \)

\State $\mathcal{T} \gets \textsc{GetMST}(N, D, E_{\text{final}})$ \Comment{Using Kruskal’s algorithm}
\State \Return $\mathcal{T}$

\end{algorithmic}
\end{algorithm*}

After constructing the initial \acp{ANN} graph, we identify its connected components using Algorithm \ref{alg:find_components} (phase 2). This algorithm performs a \ac{DFS} traversal to discover all connected components in the graph. Importantly, we first convert the directed weighted \ac{ANN} graph into an undirected weighted graph to ensure proper inter-component edge refinements during phase 3.

\begin{algorithm*}
\caption{Find Connected Components in an Undirected Graph}
\label{alg:find_components}
\begin{algorithmic}[1]
\Require Neighbors list (\(\mathcal{N}\)) representing the graph
\Ensure List of components ($C$) and the undirected graph ($\mathcal{G}$)

\State \( V \gets \emptyset \)  \Comment{Set of visited nodes}
\State \( C \gets \emptyset \)  \Comment{List of components}
\State Construct undirected graph \( \mathcal{G} \) from \( \mathcal{N} \):

\For{each \( i \in \mathcal{N} \)}
    \For{each \( j \in \mathcal{N}[i] \)}
        \State Add \( (i, j) \) and \( (j, i) \) to \( \mathcal{G} \) 
    \EndFor
\EndFor

\Function{DFS}{ $v$, $c^{\prime}$ }
    \State Initialize stack with \( v \)
    \While{stack \(\neq \emptyset\)}
        \State \( u \gets \) pop(stack)
        \If{\( u \notin V \)}
            \State \( V \gets V \cup \{u\} \) \Comment{Add the vertex to the visited set}
            \State \( c^{\prime} \gets c^{\prime}\cup \{u\} \)  \Comment{Add the vertex to the current component}
            \State Push all \( \textit{neighbor}  \in \mathcal{G}[u] \setminus V \) to stack \Comment{Add unvisited neighbors of \( u \) to the stack}
            
        \EndIf
    \EndWhile
\EndFunction

\For{each \( i \in \mathcal{G} \)}
    \If{\( i \notin V \)}
        \State \( c^{\prime} \gets \emptyset \) \Comment{New components}
        \State \textsc{DFS}( \( i, c^{\prime} \) )
        \State \( C \gets C \cup \{ c^{\prime} \} \)
    \EndIf
\EndFor

\State \Return $C$, $\mathcal{G}$

\end{algorithmic}
\end{algorithm*}

Once components are identified, we use Algorithm \ref{alg:add_random_edges} to establish initial connections between them. Rather than arbitrarily connecting components, we introduce a randomized approach that samples multiple potential edges between each component pair and selects the shortest ones. This strategy provides multiple options for inter-component connections while prioritizing shorter edges, which are more likely to appear in the true \ac{MST}.




                



\begin{algorithm*}
\caption{Add Random Edges Between Components}
\label{alg:add_random_edges}
\begin{algorithmic}[1]
\Require $X$, components $ C = \{C_1, C_2, \dots, C_t\}$, number of random edges $\lambda$
\Ensure Inter-component edges ($E$) and their corresponding components' indices ($E_C$)

\State \( E \gets \emptyset \) \Comment{Set to store newly added edges between components}
\State \( E_C \gets \emptyset \) \Comment{Set to store component index pairs corresponding to each edge}
\State \( t \gets |C| \) \Comment{Number of connected components}

\For{each \( i \in [1, t] \)}
    \For{each \( j \in [i+1, t] \)}
        \State \( \text{Candidates} \gets \emptyset \)
        \For{$\lambda^2$ times} \Comment{Generate $\lambda^2$ candidate edges}
            \State \( u \sim \text{rng}(C_i), \quad v \sim \text{rng}(C_j) \) \Comment{Randomly sample nodes}
            \State \( d \gets \|X_u - X_v\| \) \Comment{Compute distance}
            \State \( \text{Candidates} \gets \text{Candidates} \cup \{(u, v, d)\} \)
        \EndFor
        \State Sort \text{Candidates} by ascending distance
        \State Select first \( \lambda \) edges as \( \text{BestEdges} \)
        \For{each \( (u, v, d) \in \text{BestEdges} \)}
            \State \( E \gets E \cup \{(u, v, d)\} \)
            \State \( E_C \gets E_C \cup \{(i, j)\} \)
        \EndFor
    \EndFor
\EndFor

\State \Return \( E, E_C \)

\end{algorithmic}
\end{algorithm*}

The hyperparameter $\lambda$ controls the number of candidate edges considered between each component pair, allowing a trade-off between accuracy and computational efficiency.

The key innovation of our approach is the iterative refinement process described in Algorithm \ref{alg:refine_edges} (phase 3). Rather than simply accepting the initial inter-component edges, we explore local neighborhoods by looking to the neighbors of neighbors to find improved connections (closer neighbors) that reduce the total spanning tree weight.

The refinement process focuses on exploring the local neighborhoods of the current inter-component edge endpoints. For each edge $(u, v)$ connecting components $C_i$ and $C_j$, where $u \in C_i$ and $v \in C_j$, we examine all neighbors of $u$ in component $C_i$ and all neighbors of $v$ in component $C_j$ to find potentially shorter connections between the components.

We restrict this exploration to neighbors within the same component as the endpoint to avoid converting the inter-component edge $(u, v)$ into an intra-component edge where both $u$ and $v$ belong to the same component.
Such undesirable transformations could occur if the vertex under examination $u$ (or $v$) had additional connections to vertices in the neighboring component $C_j$ (or $C_i$), leading to wrong changes in the graph structure.
This local search is repeated iteratively until no further improvements are found ($|\Delta|= 0$), where $\Delta$ denotes the set of edge updates in each iteration.

\begin{algorithm*}
\caption{Refine New Edges Between Components}
\label{alg:refine_edges}
\begin{algorithmic}[1]
\Require $X$, $\mathcal{G}$,  $C$, $ E$,  $E_C$
\Ensure Refined edges \( E_{\text{refined}} \) and changes made $\Delta$

\State \( E_{\text{refined}} \gets \emptyset \), \( \Delta \gets \emptyset \) \Comment{Initialize refined edges and changes}

\For{each \( (u, v, d) \in E \) with component pair \( (C_i, C_j) \)}
    \State \( u^*, v^* \gets u, v \), \( d^* \gets d \) \Comment{Initialize best edge}
    \State \( N_u \gets \mathcal{G}[u] \cap C_i \), \( N_v \gets \mathcal{G}[v] \cap C_j \) \Comment{Neighbors in the same component}

    \For{each \( u' \in N_u \)} \Comment{Find closer vertex from the neighbors of $u$ }
        \If{\( u' \neq v \)}
            \State \( d' \gets \| X_{u'} - X_v \| \)
            \If{\( d' < d^* \)}
                \State \( u^* \gets u', d^* \gets d' \)
            \EndIf
        \EndIf
    \EndFor

    \For{each \( v' \in N_v \)} \Comment{Find closer vertex from the neighbors of $v$ }
        \If{\( v' \neq u \)}
            \State \( d' \gets \| X_{u^*} - X_{v'} \| \)
            \If{\( d' < d^* \)}
                \State \( v^* \gets v', d^* \gets d' \)
            \EndIf
        \EndIf
    \EndFor

    \If{\( (u^*, v^*) \neq (u, v) \)}
        \State \( \Delta \gets \Delta \cup \{ (u, v, d) \to (u^*, v^*, d^*) \} \) \Comment{Track change}
    \EndIf

    \State \( E_{\text{refined}} \gets E_{\text{refined}} \cup \{(u^*, v^*, d^*)\} \)
\EndFor

\State \Return \( E_{\text{refined}}, \Delta \)

\end{algorithmic}
\end{algorithm*}

The final step, following the construction of the single-component undirected graph, is to build the approximate \ac{MST} using Kruskal’s algorithm. This is achieved by applying the steps of Kruskal’s algorithm as previously described (Section \ref{sec:related}), utilizing the neighbor lists derived from the undirected graph construction along with the inter-component weighted edges obtained in Phase 3 of the \ac{FAMST} algorithm

\section{Complexity analysis}
\label{sec:complexityAnalysis}
Understanding the computational and memory efficiency of \ac{FAMST} is critical to establishing its scalability for real-world applications. This section analyzes the algorithm’s time and space complexity across its three main phases: \ac{ANN} graph construction, component connectivity, and iterative edge refinement.

\subsection{Computational complexity}
Below, we provide a detailed analysis of the computational complexity of each step of our approach.

\paragraph{\ac{ANN} graph construction}
We use the PyNNDescent library to construct the \ac{ANN} graph, though alternative algorithms exhibiting superior computational or memory efficiency could be substituted within our framework. PyNNDescent is an optimized implementation of the NN-Descent algorithm, which is widely adopted for scalable neighbor search in high-dimensional spaces.

While NN-Descent does not have a strict theoretical worst-case guarantee, 
\cite{baron2019k} suggests that the algorithm's runtime typically fits a pattern of $\mathcal{O}(k^2 d n \log n)$. 

\paragraph{Component Identification:}
After constructing the \ac{ANN} graph, our algorithm identifies connected components using a \ac{DFS} approach:
\begin{itemize}
    \item \textbf{Graph representation}: converting the \ac{ANN} list representation to an undirected graph requires $\mathcal{O}(kn)$ operations
    \item \textbf{\ac{DFS} Traversal}: for each vertex in the graph, DFS explores all adjacent edges; With $n$ vertices and approximately $k \times n$ edges, this yields a complexity of $\mathcal{O}(kn)$
\end{itemize}

The total complexity for component identification is therefore $\mathcal{O}(kn)$, which scales linearly with the dataset size given a fixed $k$.

\paragraph{Random edge addition:}
To connect components, our algorithm evaluates multiple candidate connections between each pair of disconnected components:
\begin{itemize}
    \item For $t$ components, there are   $\frac{t(t-1)}{2} = \mathcal{O}(t^2)$ unique component pairs.
    \item For each pair, the algorithm randomly samples $\lambda^2$ candidate edges by selecting node pairs from the two components then selects the top $\lambda$ shortest edges (by distance) and adds them to the inter-component edge set.
    \item This results in $\mathcal{O}(\lambda^2)$ distance computations and an additional $\mathcal{O}(\lambda^2 \log \lambda^2) = \mathcal{O}(2\lambda^2 \log \lambda) = \mathcal{O}(\lambda^2 \log \lambda)$ cost for sorting per component pair. However, $\lambda$ is a small constant, so this step remains efficient and scales well with the number of components.
\end{itemize}

In practice, the number of components $t$ is typically much smaller than $n$ and decreases rapidly as $k$ increases. For most datasets, $t \ll n$ makes this stage practically efficient.

\paragraph{Edge refinement:}

\begin{itemize}
    \item Neighborhood exploration $\mathcal{O}(t \lambda k)$: for each of the  $\mathcal{O}(t \lambda )$ inter-component edges, we explore $k$ neighbors in each component. Each exploration requires $\mathcal{O}(1)$ distance calculations.
 
    \item Iteration Count $\mathcal{O}(r)$: The number of refinement iterations $r$ is typically small and does not scale with the dataset size $n$, but it may depend on the neighborhood size $k$ and the number of inter-component edges $\lambda$, as these parameters influence the number and quality of initial inter-component connections that the refinement process seeks to improve.
\end{itemize}

The total complexity for the refinement process is $\mathcal{O}(r t \lambda k)$. Since $t \ll n$ in practice and $r$ is typically small and could be set as a small constant, this stage also maintains near-linear scaling with dataset size.

\paragraph{\ac{MST} extraction using Kruskal’s algorithm}
Once the graph is connected, we apply Kruskal’s algorithm which has time complexity equal to $\mathbf{O}(E \log E)$ where $E$ is the number of edges in the final graph. In \ac{FAMST}, the final graph has $E= \mathcal{O}(kn+t\lambda)$, so Kruskal’s step runs in:
$$\mathcal{O}\left((kn + t\lambda)\log(kn + t\lambda)\right)$$

\paragraph{Overall time complexity:}
Combining all steps, the time complexity of \ac{FAMST} is
\[
\mathcal{O}(k^2dn\log n) 
+ \mathcal{O}(kn) 
+ \mathcal{O}(t^2\lambda^2 \log \lambda^2) 
+ \mathcal{O}(rt\lambda k) 
+ \mathcal{O}\left((kn + t\lambda)\log(kn + t\lambda)\right),
\]
For practical settings where $t \ll n$ and $k$, $\lambda$, and $r$ are small constants, the overall time complexity simplifies to:
\[
\mathcal{O}(dn \log n),
\]
While the time complexity of our approach is often dominated by the cost of \ac{ANN} computation, this is not always the case. In particular, under \emph{unrecommended settings}—such as using a small neighborhood size $k$ together with a large number of inter-component edges $\lambda$—the refinement phase can become the computational bottleneck due to the large number of inter-component edges that must be evaluated in each iteration (see Section~\ref{sec:hyperparameters}).

\subsection{Space complexity}
\label{sec:space_complexity}

The space complexity arises from storing the dataset, the \ac{ANN} graph, component metadata, and inter-component connections. Below, we analyze each contribution:

\begin{itemize}
\item \textbf{Data storage:} The input dataset $X \in \mathbb{R}^{n \times d}$ requires $\mathcal{O}(dn)$ memory.

\item \textbf{\ac{ANN} graph:} For each of the $n$ data points, we store the indices of its $k$ nearest neighbors as well as their corresponding distances, requiring $k$ entries for both. This results in a total space cost of $2kn$. However, in asymptotic analysis, constant factors are omitted, yielding a space complexity of $\mathcal{O}(kn)$.


\item \textbf{Component labels:} Maintaining the component membership of each node requires $\mathcal{O}(n)$ space.

\item \textbf{Inter-component edges:} A total of $\mathcal{O}(t\lambda^2)$ inter-component edges are maintained during the refinement phase, where $t$ is the number of connected components. Since $t \ll n$ in practice, this term is negligible.
\end{itemize}
Combining all terms (omitting the negligible ones), the overall space complexity is:
\[
\mathcal{O}(dn + kn),
\]
given that $k \ll d$ for high-dimensional applications, the dominant term is $\mathcal{O}{(dn)}$, meaning the algorithm scales linearly with both the number of points and their dimensionality

\section{Experimental evaluation}
\label{sec:experiments}
All experiments were conducted on a desktop machine equipped with an AMD Ryzen 5 7600X (6-core, 12-thread, 5.45 GHz max boost), 64 GB DDR5 RAM, and Ubuntu 22.04 LTS operating system. To ensure statistical reliability, all reported metrics represent the mean values computed over $10$ runs; each run was initialized with a different random seed to account for variability in the \ac{ANN} graph construction and random inter-component edges sampling.

\subsection{Datasets}
To verify the effectiveness of \ac{FAMST}, we tested it across diverse datasets as outlined in Table \ref{tab:datasets_and_results}.
The experimental datasets included Speech, MNIST, and Shuttle from the Outlier Detection Data Sets\cprotect\footnote{\url{https://github.com/Minqi824/ADBench}}; Fashion-MNIST\footnote{\url{https://github.com/zalandoresearch/fashion-mnist}} (referred to as F-MINST below) which provides clothing item images; CelebA introduced in~\cite{liu2015deep} and available online\cprotect\footnote{\url{https://paperswithcode.com/dataset/celeba}}; 
several real-world datasets (Miss America, House, Europe) and synthetic datasets (Unbalanced, Birch) obtained from a public repository\cprotect\footnote{\url{https://cs.joensuu.fi/sipu/datasets/}}; Corel, Shape, and Audio datasets utilized in prior research~\cite{dong2011efficient, sieranoja2018constructing} and accessible online\cprotect\footnote{\url{https://code.google.com/archive/p/nndes/downloads}}; and Blobs datasets with varying $d$ and $n$ created using the \emph{make\_blobs} function from \emph{Sklearn}\cprotect\footnote{\url{https://scikit-learn.org/stable/modules/generated/sklearn.datasets.make_blobs.html}}.

\subsection{Metrics}
To evaluate the performance of \ac{FAMST}, we employed two primary metrics that assess both computational efficiency and approximation quality: 

\paragraph{Computational efficiency}
We measure computational efficiency by recording the total elapsed time in seconds, capturing the total processing time required for \ac{MST} construction. This metric allows for direct comparison with baseline algorithms across datasets of varying scale and dimensionality. 

\paragraph{Approximation quality}
To quantify the approximation quality, we employ the relative error metric, which measures the deviation of the approximate \ac{MST} weight from the exact \ac{MST} weight as defined in \eqref{eq:relativeError}. The relative error provides a scale-invariant, normalized measure of accuracy, allowing for fair comparisons across datasets of varying size and dimensionality.

\begin{equation}
 \text{Relative error} = \frac{W_{\text{approx}} - W_{\text{exact}}}{W_{\text{exact}}} 
\label{eq:relativeError}
\end{equation}
where $W_{\text{approx}}$ represents the total edge weight of the \ac{MST} constructed by \ac{FAMST}, and $W_{\text{exact}}$ denotes the total edge weight of the exact \ac{MST} obtained using \ac{EMST} method implementation from the \textbf{mlpack} library\cprotect\footnote{\url{https://www.mlpack.org/}}.




\subsection{Results}
\label{results}
Table \ref{tab:datasets_and_results} presents the performance comparison between our proposed algorithm (\ac{FAMST}) and other methods from the literature across the 14 datasets. The empirical results demonstrate several significant findings:

\ac{FAMST} achieves remarkably low approximation errors across all tested datasets, with a mean relative error of 0.44\% and a median of 0.07\%.
This represents a substantial improvement over the \ac{MFC} algorithm, which yields a mean relative error of 4.5\% (the second best after \ac{FAMST}). Notably, on datasets such as MNIST and Shuttle, \ac{FAMST} achieves relative errors of 0.049\% and 0.045\% respectively, indicating near-optimal solutions.

The computational efficiency of \ac{FAMST} is particularly evident in high-dimensional and large-scale datasets. For instance, on the F-MNIST dataset (n=70,000, d=784), \ac{FAMST} completes in $4.39$ seconds compared to $4172.99$ seconds for the exact \ac{EMST} algorithm (a 950× speedup while maintaining a relative error of merely 0.071\%). This reduction in computational time without significant sacrifice in solution quality highlights the practical utility of our approach for large-scale applications.

In a few datasets, such as Shuttle, Europe, Unbalanced, and Birch1, \ac{FAMST} does not exhibit a performance advantage over \ac{EMST}. These datasets are characterized by low dimensionality (2–9 dimensions) and relatively small sample sizes, conditions under which \ac{EMST} algorithm can be highly efficient due to its optimized implementation. In contrast, \ac{FAMST} is primarily designed to scale effectively in large-scale, high-dimensional scenarios, where traditional \ac{MST} methods become computationally prohibitive.

\begin{table*}[ht]
\centering
\small
\begin{tabular}{@{}l l l l l l l l l l l@{}}
    \toprule
    \multirow{2}{*}{\textbf{Dataset}} & \multirow{2}{*}{\textbf{Type}} & \multirow{2}{*}{\textbf{$d$}} & \multirow{2}{*}{\textbf{$n$}} & \multicolumn{3}{c}{\textbf{Rel. Error (\%)}} & \multicolumn{4}{c}{\textbf{Time (s)}} \\
    \cmidrule(lr){5-7} \cmidrule(lr){8-11}
     &  &  &  &\textbf{\scriptsize MFC} & \textbf{\scriptsize FISHDBC}  & \textbf{\scriptsize\ac{FAMST}}  & \textbf{\scriptsize\ac{EMST}} & \textbf{\scriptsize MFC} & \textbf{\scriptsize FISHDBC} & \textbf{\scriptsize\ac{FAMST}} \\
    \midrule
    Speech             & \multirow{12}{*}{Real}      & 400     & 3,686   &  \textbf{0.291} & 20.480 & 2.666   & 23.67        &  157.57 & 7.52  & \textbf{0.19} \\
    Miss America\!\!\!       &                             & 16      & 6,480   &  6.444 & 16.396 & \textbf{0.129}   & 1.9          &   2.14  & 13.28 & \textbf{0.38} \\
    MNIST              &                             & 100     & 7,603   &  9.042 & 18.001 & \textbf{0.049}   & 4.91         &   16.91 & 14.60 & \textbf{0.48} \\
    Shape              &                             & 544     & 28,775  &  7.006 & 38.089 & \textbf{0.104}   & 123.18       &  1518.25 & 77.11 & \textbf{1.51} \\
    House              &                             & 3       & 34,112  &  5.104 & 50.255 & \textbf{0.021}   & \textbf{0.2} &  5.19    & 40.68 & 0.67 \\
    Shuttle            &                             & 9       & 49,097  &  1.693 & 106.838 & \textbf{0.045}   & \textbf{0.61}& 125.04   & 85.82 &1.34 \\
    Audio              &                             & 192     & 54,387  &  6.199 & 14.105 & \textbf{0.135}   & 394.4        &  513.23  & 240.37 &\textbf{2.84} \\
    F-MNIST$^*$        &                             & 784     & 30,000  &  3.601 & 130.874 & \textbf{0.396}   & 778.30       &  2875.60 &102.85 &\textbf{1.86} \\
    F-MNIST            &                             & 784     & 70,000  &  3.255 & 141.117 & \textbf{0.071}   & 4172.99      &  9805.74 & 309.57 &\textbf{4.39} \\
    Europe             &                             & 2       & 169,308 &  \multicolumn{1}{c}{---}  & 92.412 & \textbf{0.000 }  & \textbf{0.71}&  \multicolumn{1}{c}{---}    & 291.53 &3.74 \\
    CelebA             &                             & 39      & 202,599 &  7.660 & 96.123 & \textbf{4.124}   & 13.47        &  984.69  & 613.88 & \textbf{9.25} \\
    Corel              &                             & 14      & 662,317 &  3.881 & \multicolumn{1}{c}{---} & \textbf{0.026}   & 219.41       &  1592.55 & \multicolumn{1}{c}{---} &\textbf{38.45} \\
    \midrule
    Unbalanced         & \multirow{3}{*}{Synth.}  & 2       & 6,500   &  4.436 & 82.644 & \textbf{0.035}   & \textbf{0.02}&  0.68    & 4.69 & 0.20 \\
    Birch1             &                             & 2       & $10^5$  &  2.903 & 94.932 & \textbf{0.000}   & \textbf{0.38}& 11.83    & 102.49 & 2.17 \\
    Blobs              &                             & 600     & 50,000  &  2.272  & 2.923 & \textbf{1.220}  & 3209.89   & 2903.83  & 194.37 &\textbf{3.04} \\
    \bottomrule
\end{tabular}
\caption{Datasets description and performance comparison between \ac{EMST} and \ac{FAMST} with $k = 10$ and number of random edges $\lambda = 5$. F-MNIST$^*$ is a random subset ($n=30,000$) of F-MNIST.} The (---) denotes a memory error.  
\label{tab:datasets_and_results}
\end{table*}

\subsection{Dimensionality and performance}
\label{sec:Dimensionality}
To evaluate the scalability of \ac{FAMST} with respect to increasing dimensionality and dataset size, we conduct experiments using synthetic datasets with varying feature dimensions and sample counts. For each experiment, we generate datasets with fixed sample sizes ($n$) and vary the dimensionality ($d$), measuring the total execution time of the algorithm. Each configuration is run multiple times to ensure consistency, and we report average computation times.

\begin{figure*}
    \centering
    \includegraphics[width=0.75\linewidth]{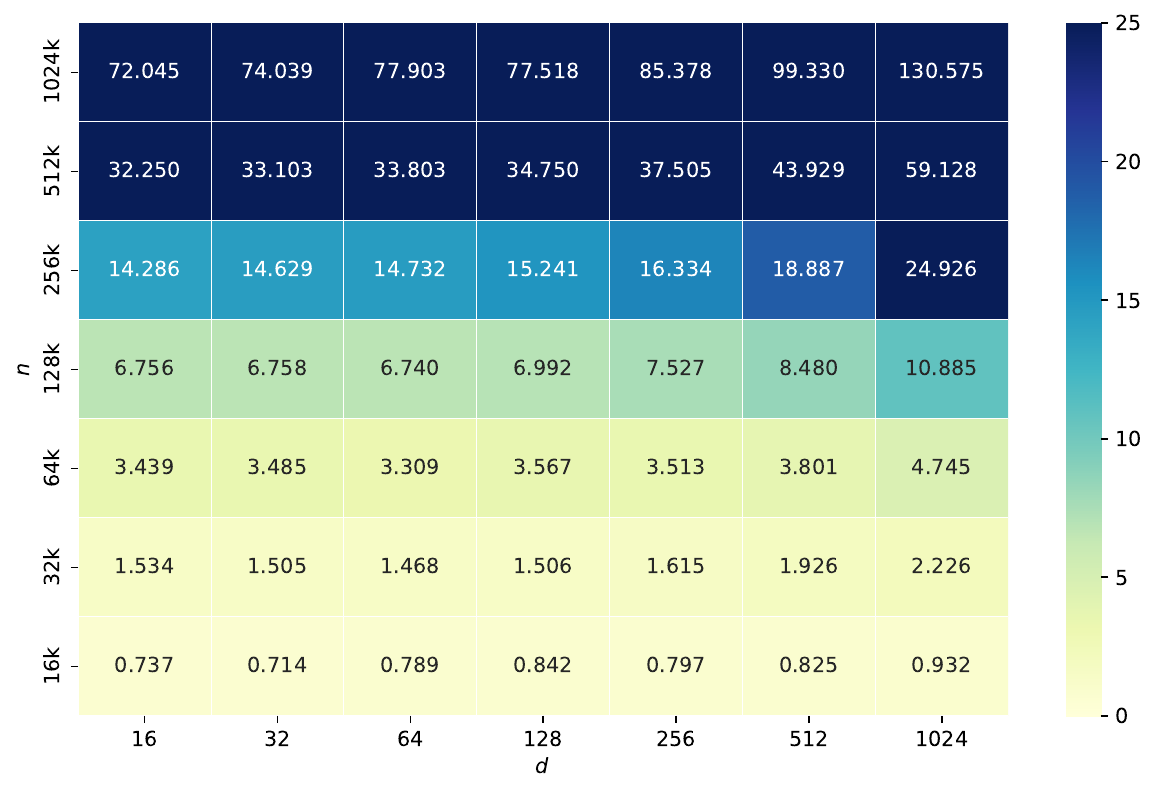}
    \caption{The impact of $d$ on the \ac{FAMST} running time.}
    \label{fig:heatmap}
\end{figure*}

Figure~\ref{fig:heatmap} illustrates the runtime behavior of \ac{FAMST} across various dimensionalities and dataset sizes. The results confirm that \ac{FAMST} scales favorably: execution time increases approximately linearly with the number of points $n$, and only moderately with the dimensionality $d$. This behavior is in line with our theoretical time complexity of $\mathcal{O}(dn \log n)$.

\begin{figure*}
     \centering
     \begin{subfigure}[b]{0.49\textwidth}
         \centering
         \includegraphics[width=\textwidth]{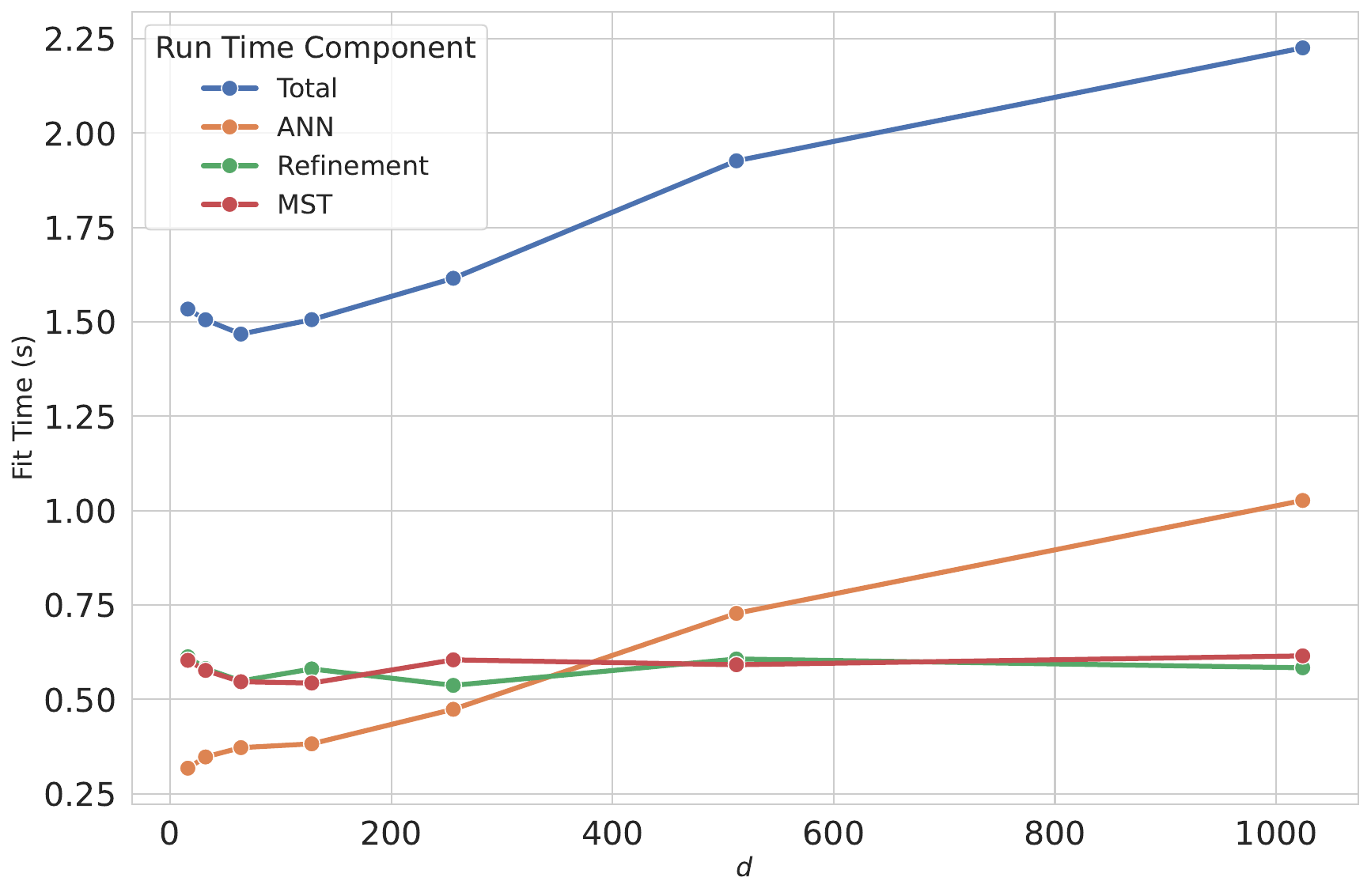}
         \caption{$n=32,000$}
     \end{subfigure}
     \hfill
     \begin{subfigure}[b]{0.49\textwidth}
         \centering
         \includegraphics[width=\textwidth]{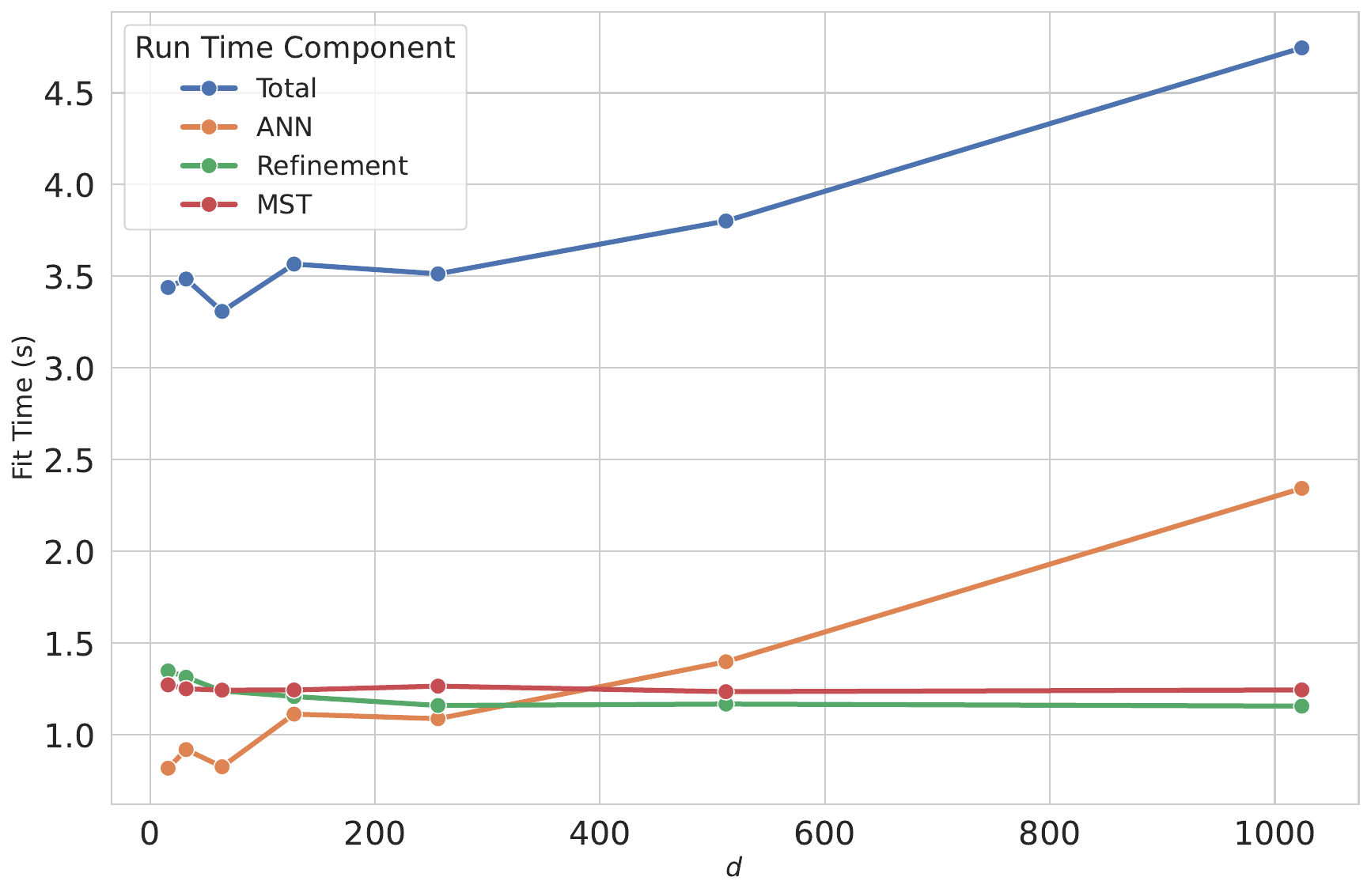}
         \caption{$n=64,000$}
     \end{subfigure}
     \hfill
     \begin{subfigure}[b]{0.49\textwidth}
         \centering
         \includegraphics[width=\textwidth]{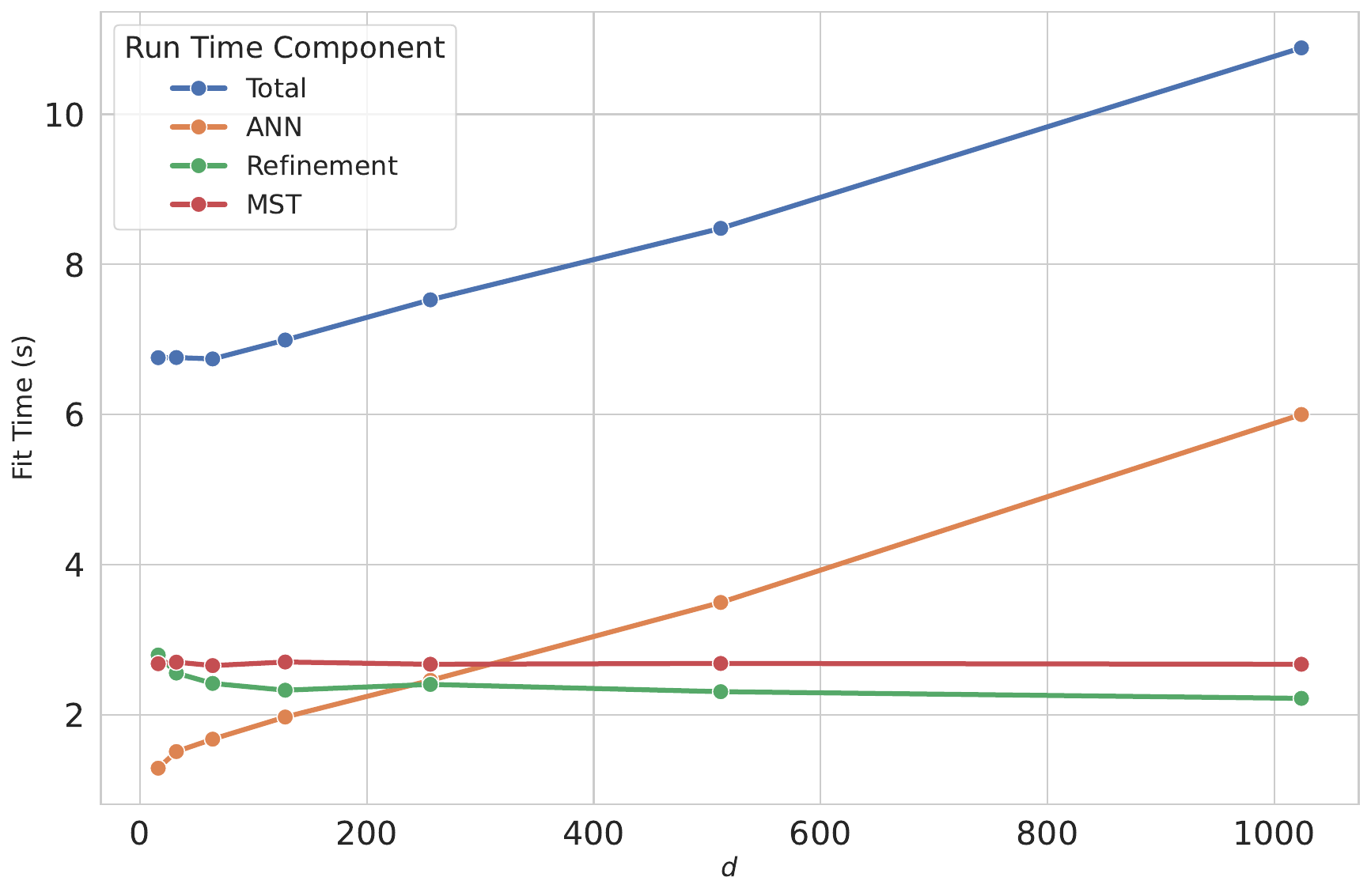}
         \caption{$n=128,000$}
     \end{subfigure}
     \hfill
     \begin{subfigure}[b]{0.49\textwidth}
         \centering
         \includegraphics[width=\textwidth]{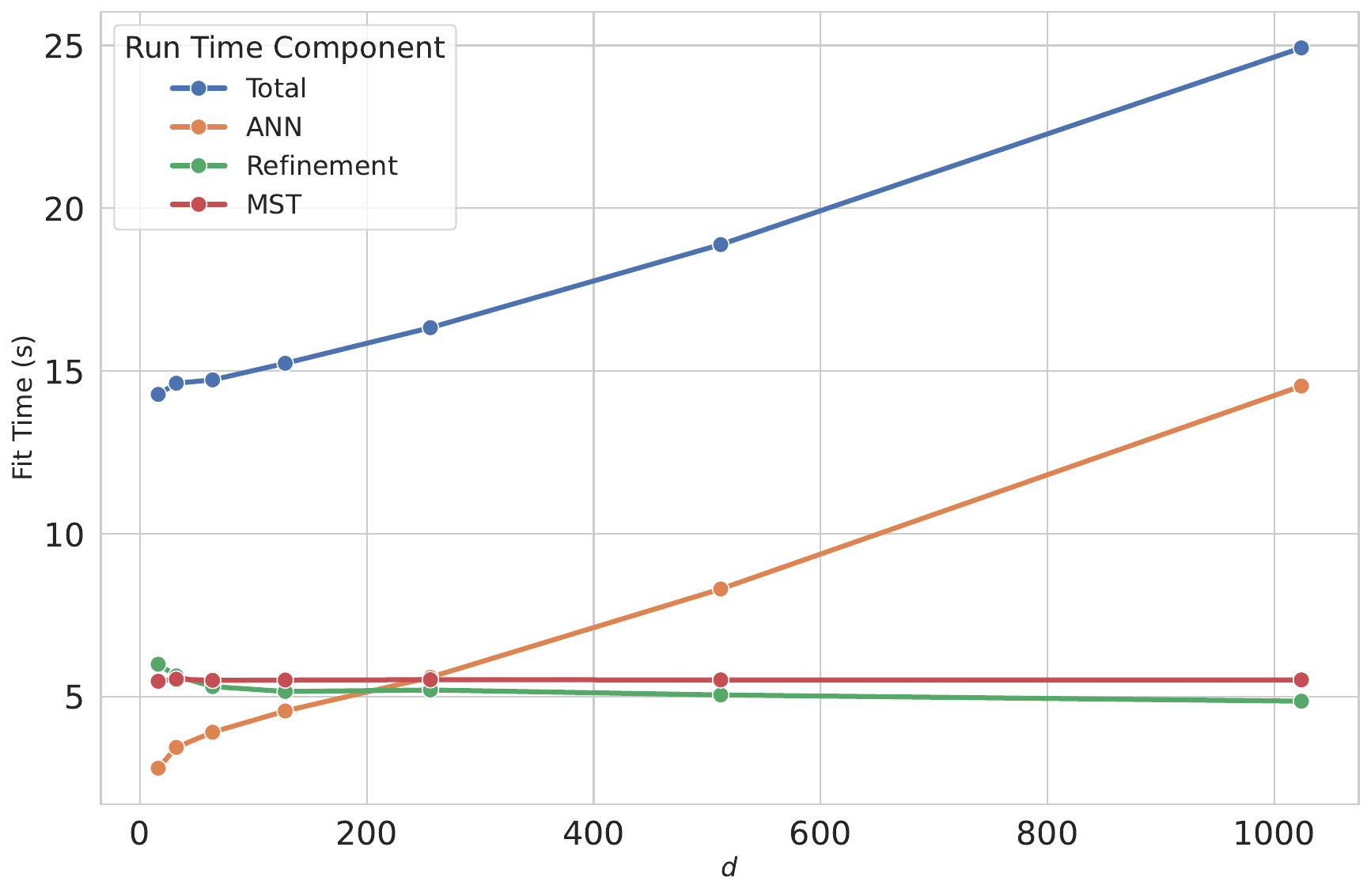}
         \caption{$n=356,000$}
     \end{subfigure}
     \hfill
     \begin{subfigure}[b]{0.49\textwidth}
         \centering
         \includegraphics[width=\textwidth]{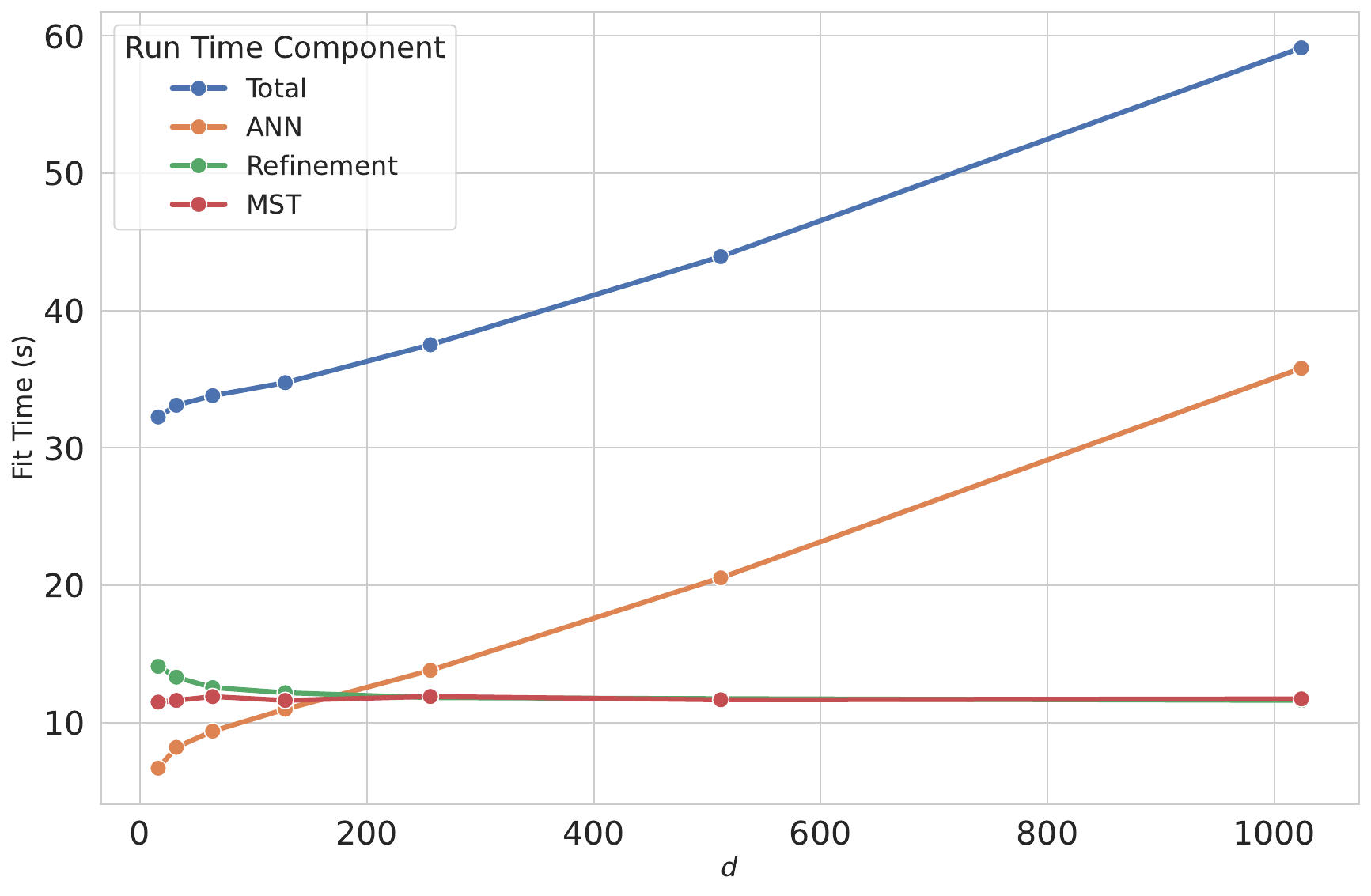}
         \caption{$n=512,000$}
     \end{subfigure}
     \hfill
     \begin{subfigure}[b]{0.49\textwidth}
         \centering
         \includegraphics[width=\textwidth]{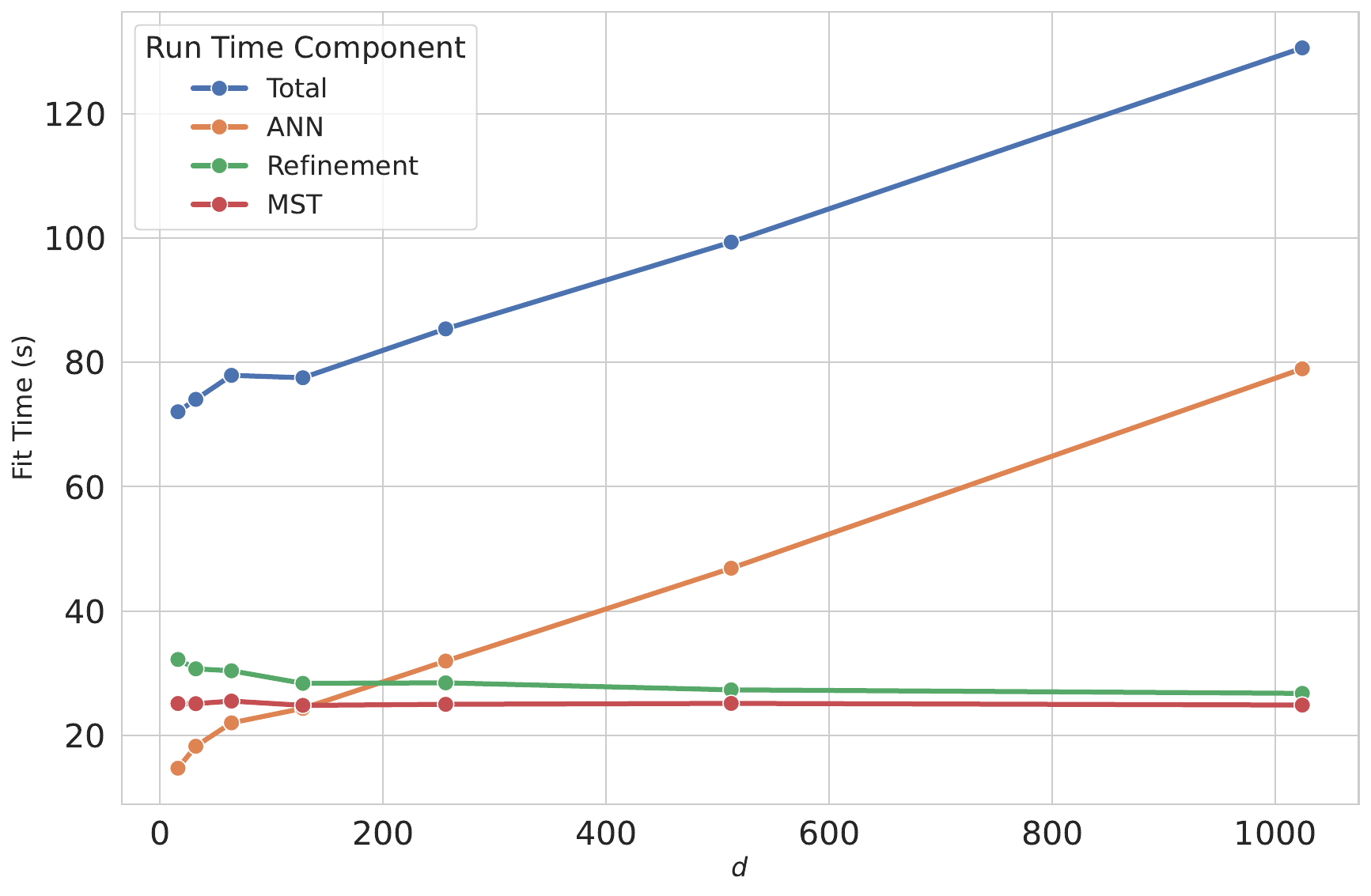}
         \caption{$n=1,024,000$}
     \end{subfigure}
        \caption{Computational complexity analysis: Phase-by-phase \ac{FAMST} execution time versus dimensionality}
        \label{fig:effectOfdLinePlot}
\end{figure*}

To better understand how each phase of the algorithm contributes to total runtime, we decompose the execution time into three components: \ac{ANN} graph construction, edge refinement, and \ac{MST} extraction. As shown in Figure~\ref{fig:effectOfdLinePlot}, \ac{ANN} graph construction consistently dominates the total cost, particularly for large values of $n$ or $d$. In contrast, the refinement and \ac{MST} phases exhibit stable and comparatively negligible overhead, even for high-dimensional datasets.

These findings emphasize the scalability of \ac{FAMST}. The refinement process, in particular, demonstrates strong efficiency as its cost remains effectively low regardless of input dimensionality or size.

\section{Hyperparameters analysis}
\label{sec:hyperparameters}

The performance of \ac{FAMST} depends significantly on two hyperparameters: the number of nearest neighbors $k$, and the number of inter-component connections $\lambda$. We investigate their impact on both accuracy and runtime across diverse datasets.

\begin{figure*}
     \centering
     \begin{subfigure}[b]{0.49\textwidth}
         \centering
         \includegraphics[width=\textwidth]{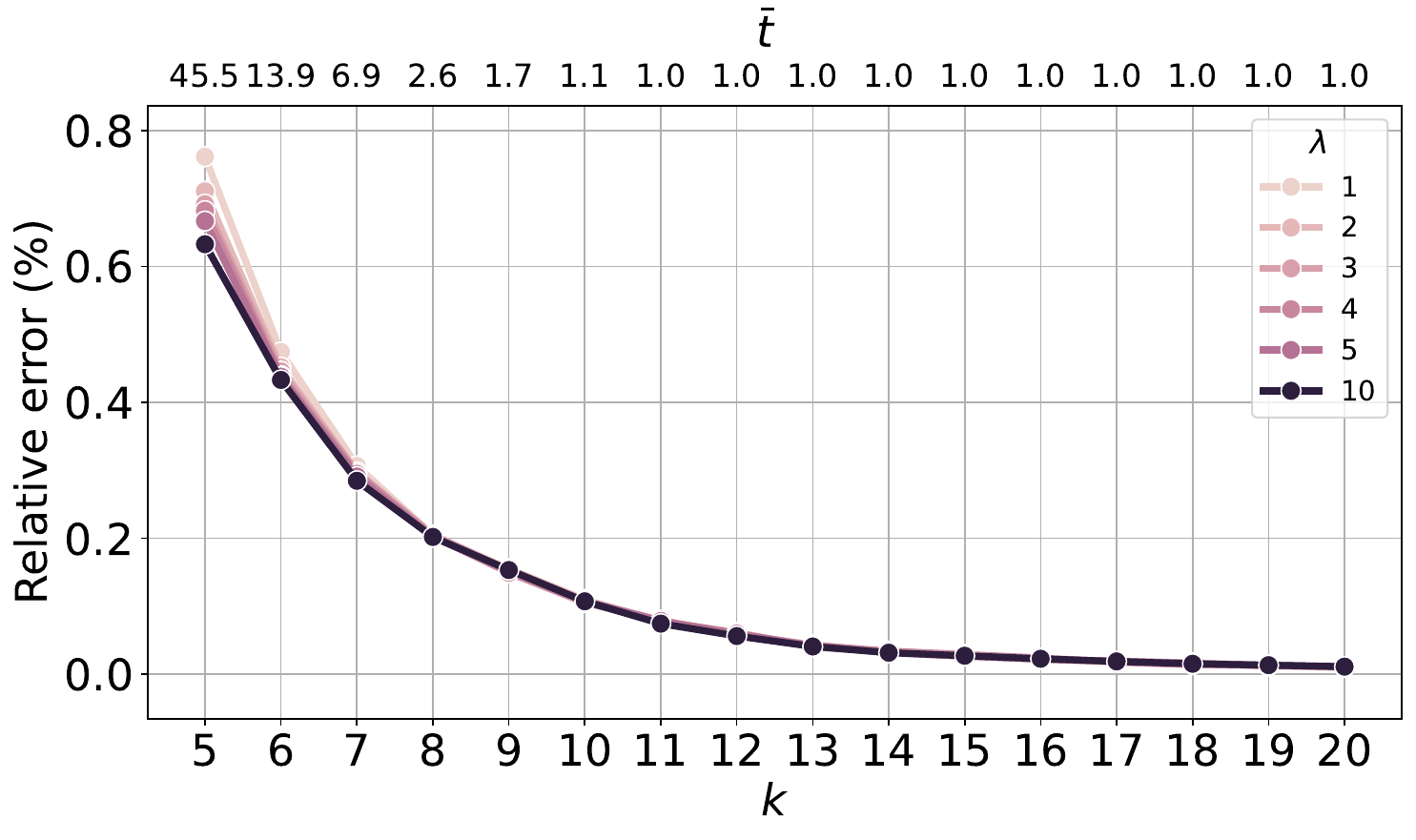}
         \caption{Shape}
     \end{subfigure}
     \hfill
     \begin{subfigure}[b]{0.49\textwidth}
         \centering
         \includegraphics[width=\textwidth]{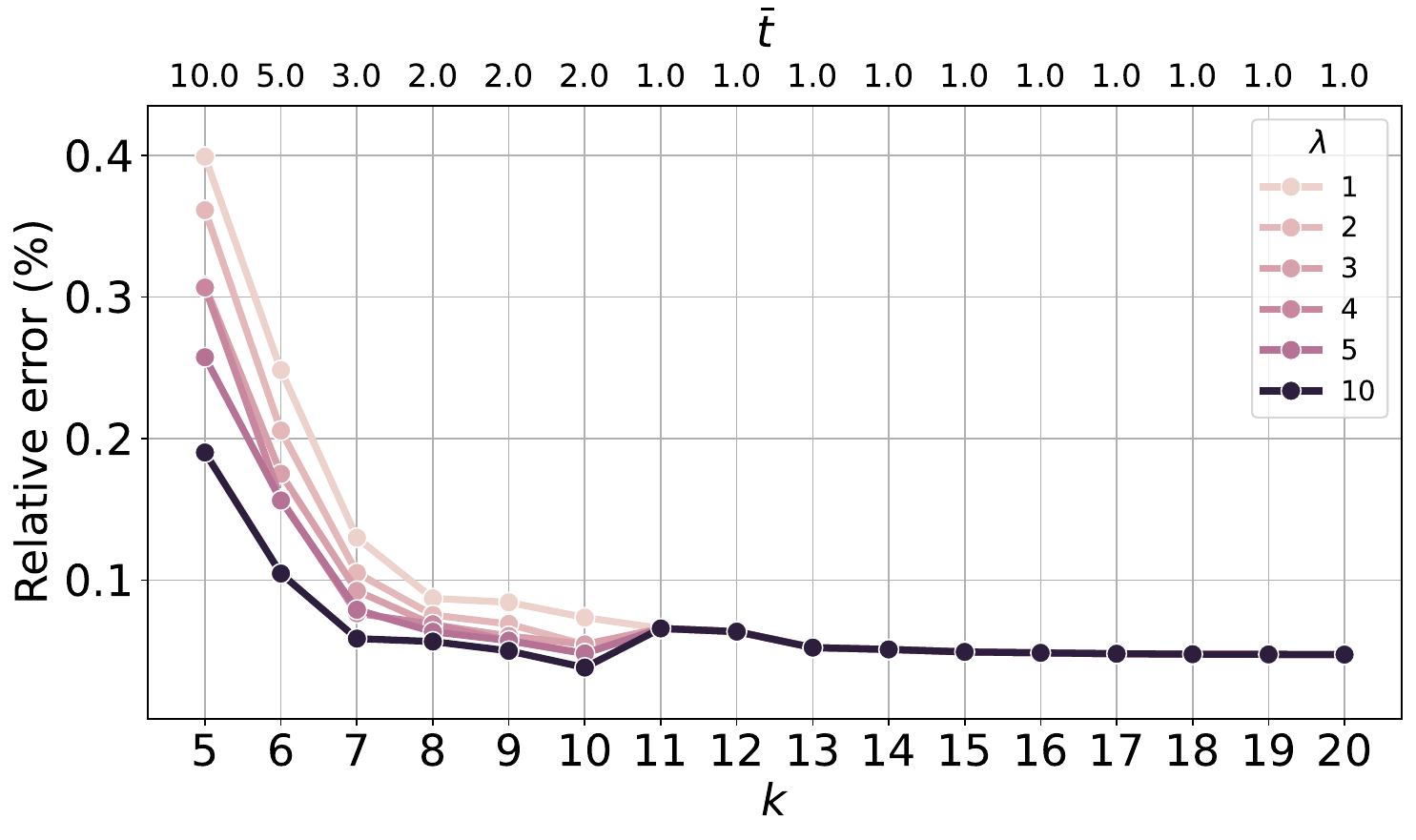}
         \caption{Shuttle}
     \end{subfigure}
     \hfill
     \begin{subfigure}[b]{0.49\textwidth}
         \centering
         \includegraphics[width=\textwidth]{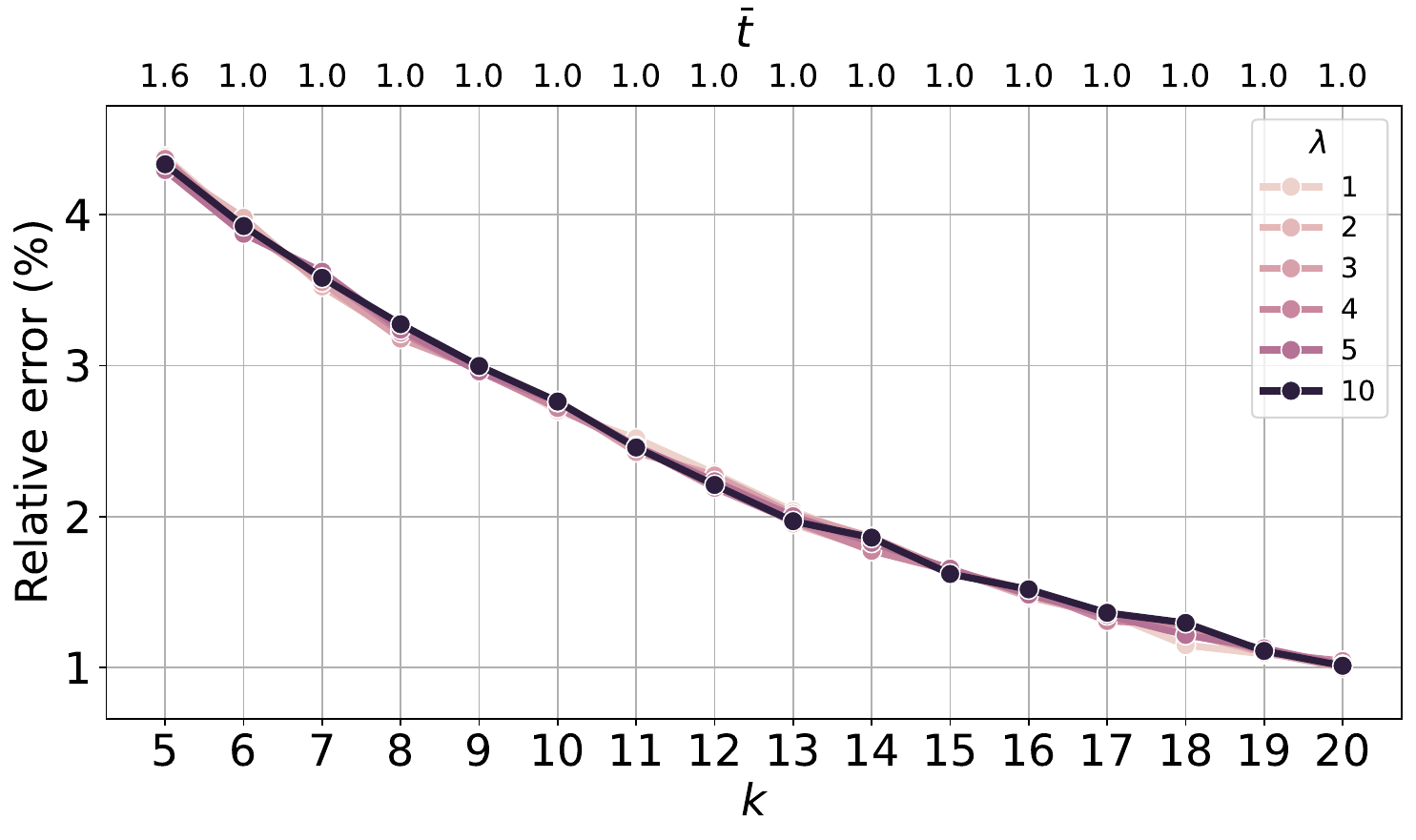}
         \caption{Speech}
          \label{fig:speech_error_rate}
     \end{subfigure}
     \hfill
     \begin{subfigure}[b]{0.49\textwidth}
         \centering
         \includegraphics[width=\textwidth]{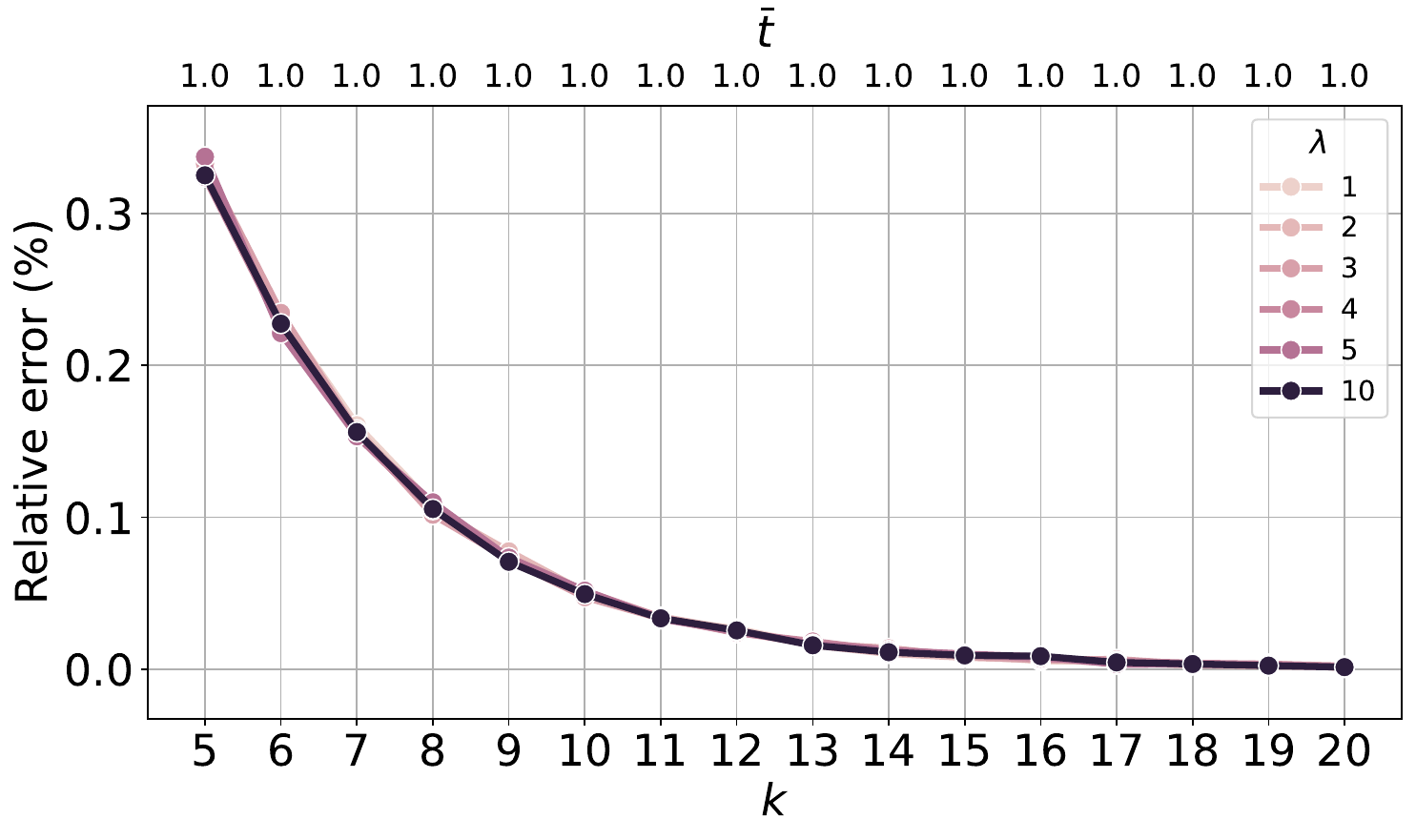}
         \caption{MNIST}
     \end{subfigure}
     \hfill
     \begin{subfigure}[b]{0.49\textwidth}
         \centering
         \includegraphics[width=\textwidth]{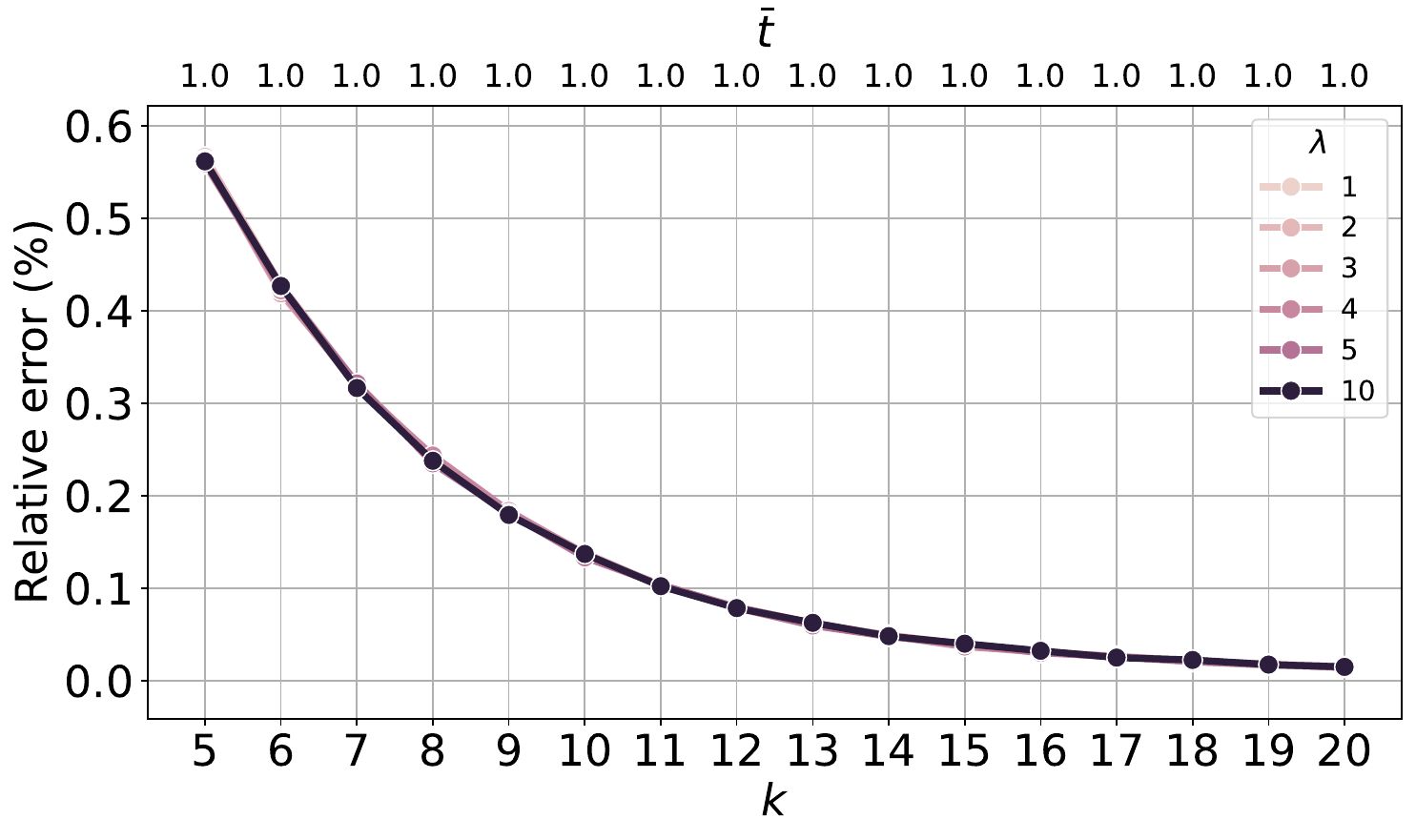}
         \caption{Audio}
     \end{subfigure}
     \hfill
     \begin{subfigure}[b]{0.49\textwidth}
         \centering
         \includegraphics[width=\textwidth]{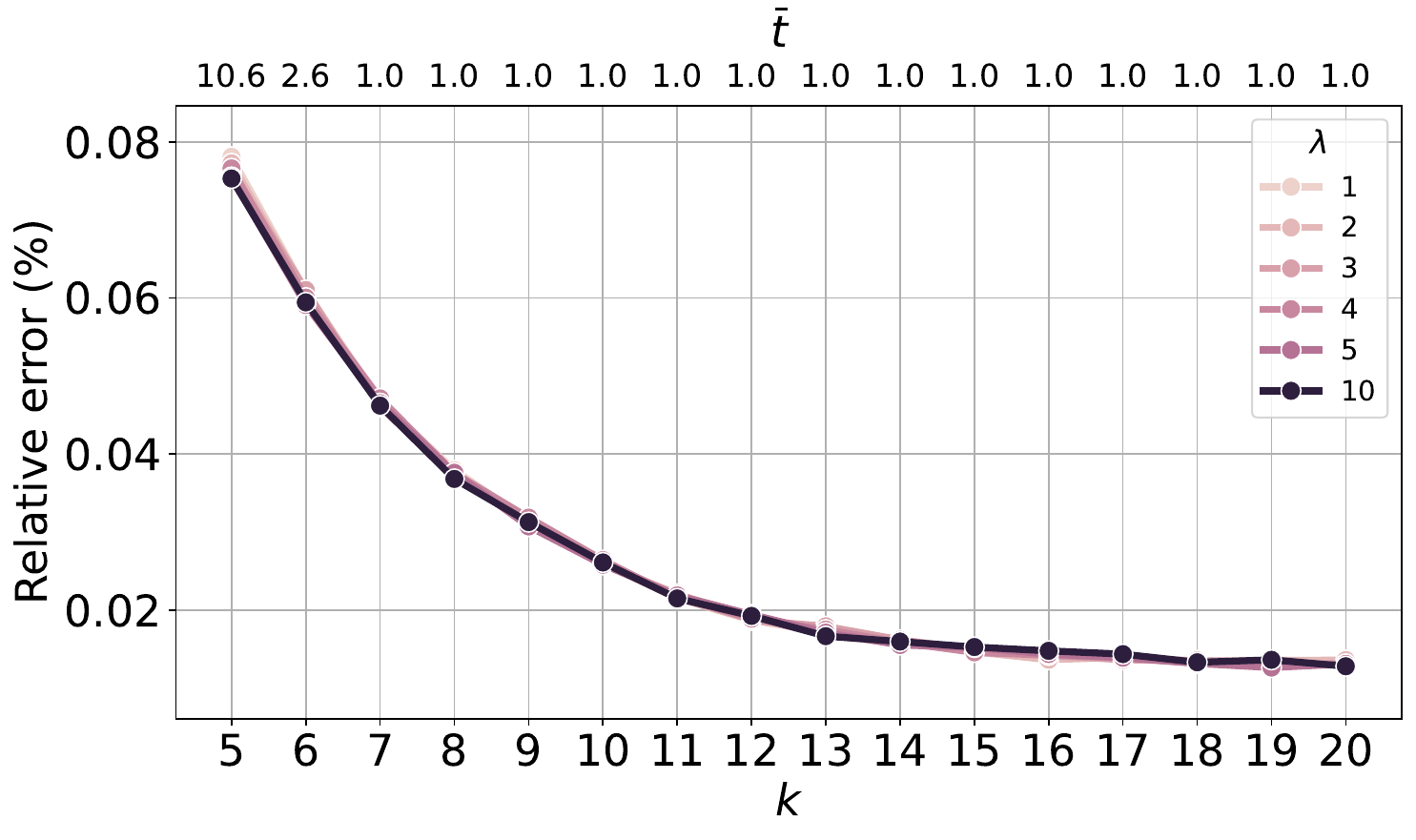}
         \caption{Corel}
     \end{subfigure}
     \hfill
     \begin{subfigure}[b]{0.49\textwidth}
         \centering
         \includegraphics[width=\textwidth]{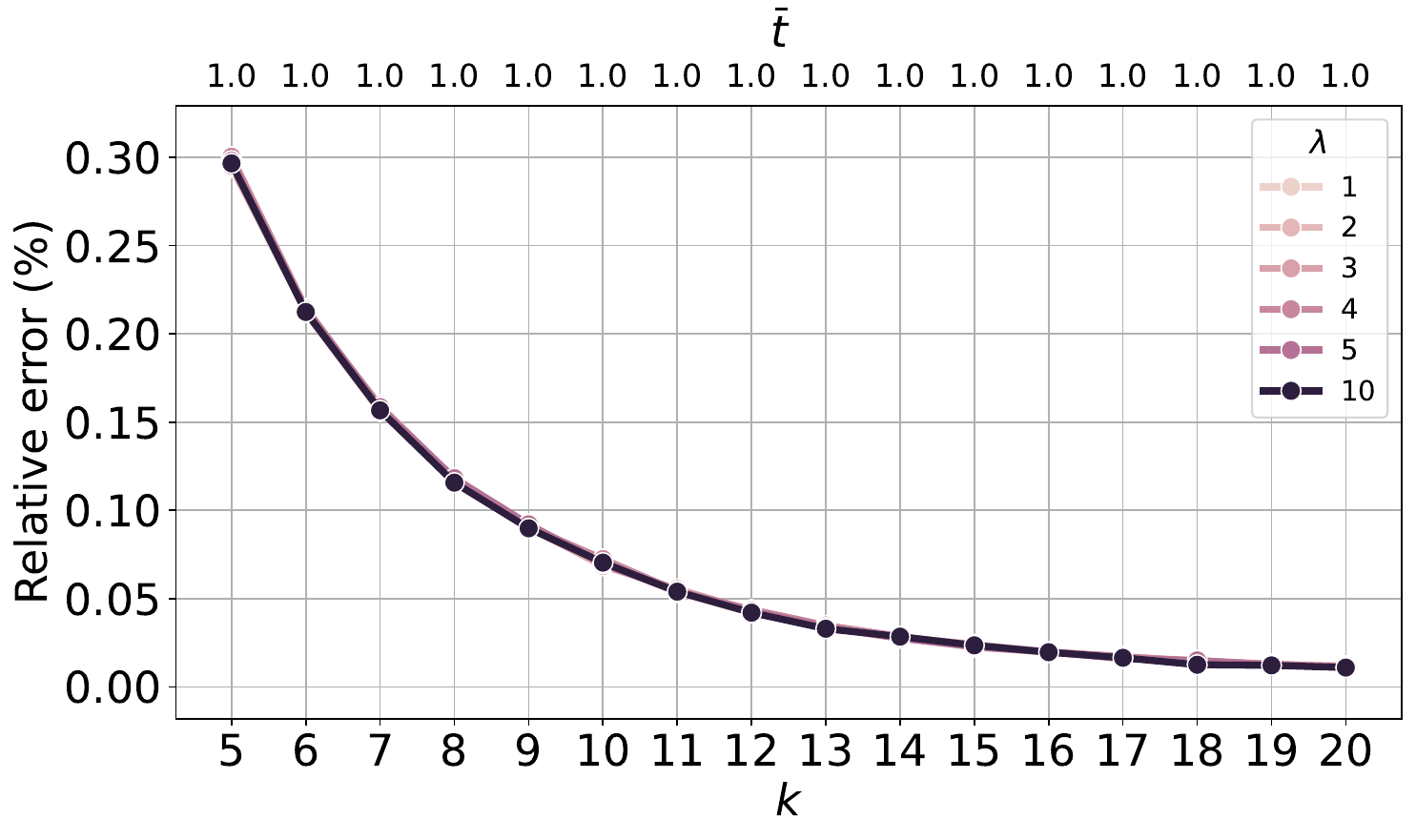}
         \caption{F-MNIST}
     \end{subfigure}
     \hfill
     \begin{subfigure}[b]{0.49\textwidth}
         \centering
         \includegraphics[width=\textwidth]{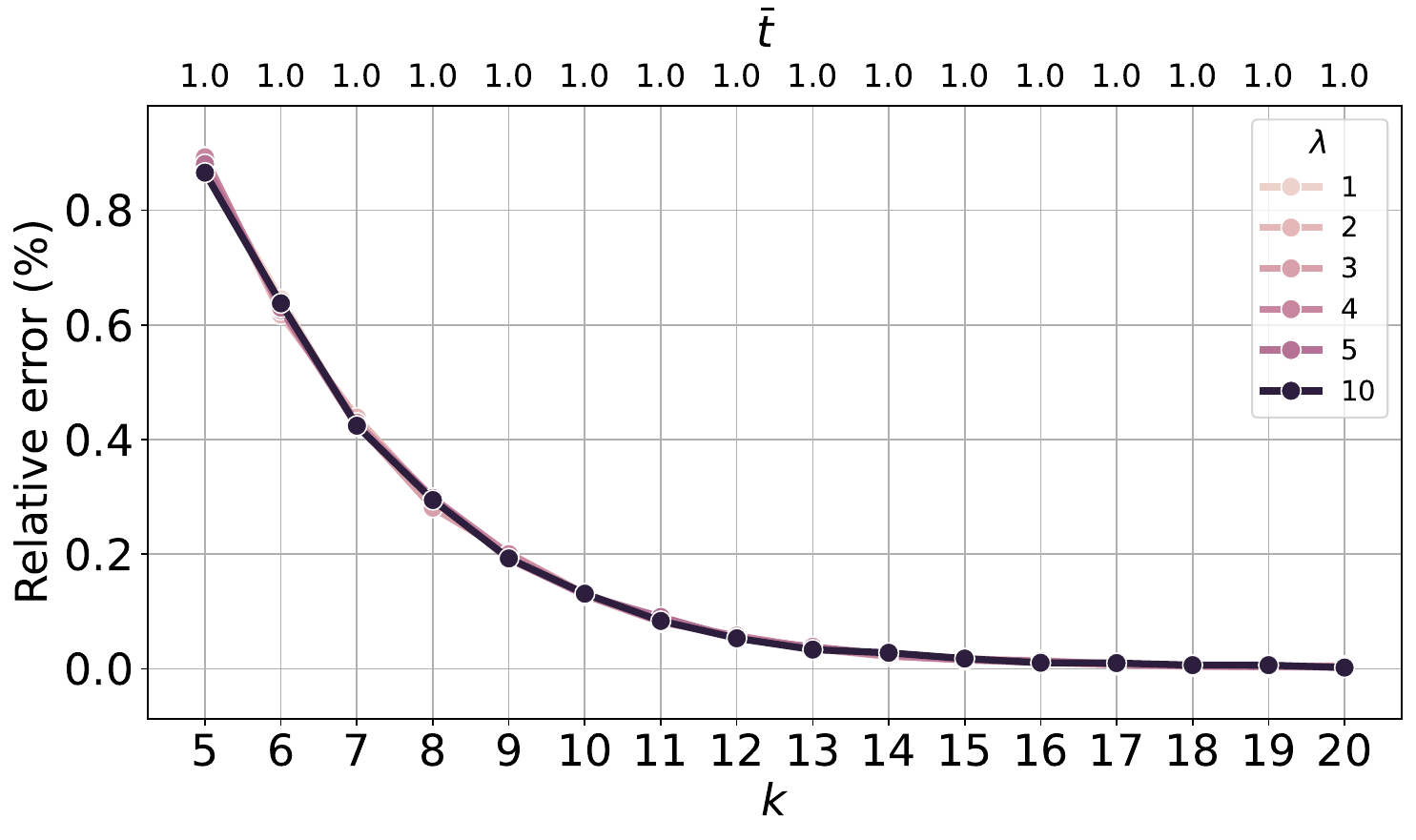}
         \caption{Miss America}
     \end{subfigure}
        \caption{Relative error (\%) for different datasets using different $k$ and $\lambda$ values; $\bar{t}$ is the average of $t$ (number of disconnected components in the \ac{ANN} graph) over different runs}
        \label{fig:errorRateLineGraph}
\end{figure*}

\paragraph{Effect on approximation error}
As illustrated in Figure \ref{fig:errorRateLineGraph}, the hyperparameter $k$ has a dominant and consistent effect on approximation error across all tested datasets. Increasing $k$ from 5 to 20 substantially reduces the error rate for example, in the Speech dataset (Figure \ref{fig:speech_error_rate}), the error decreases from approximately 4.5\% to 1.5\%. 
In contrast, $\lambda$ demonstrates minimal impact on approximation error for most datasets. The error curves for different $\lambda$ values (1, 2, 3, 4, 5, and 10) cluster closely together at each $k$ value. This suggests that even a small number of random inter-component edges ($\lambda$ = 1 or 2) is often sufficient to achieve good approximation quality.

This behavior aligns with the theoretical foundation of our approach - the $k$ hyperparameter directly determines how well the local neighborhood structure is captured in the initial \ac{ANN} graph, which has a fundamental impact on the \ac{MST} approximation quality. 

\begin{figure*}
     \centering
     \begin{subfigure}[b]{0.49\textwidth}
         \centering
         \includegraphics[width=\textwidth]{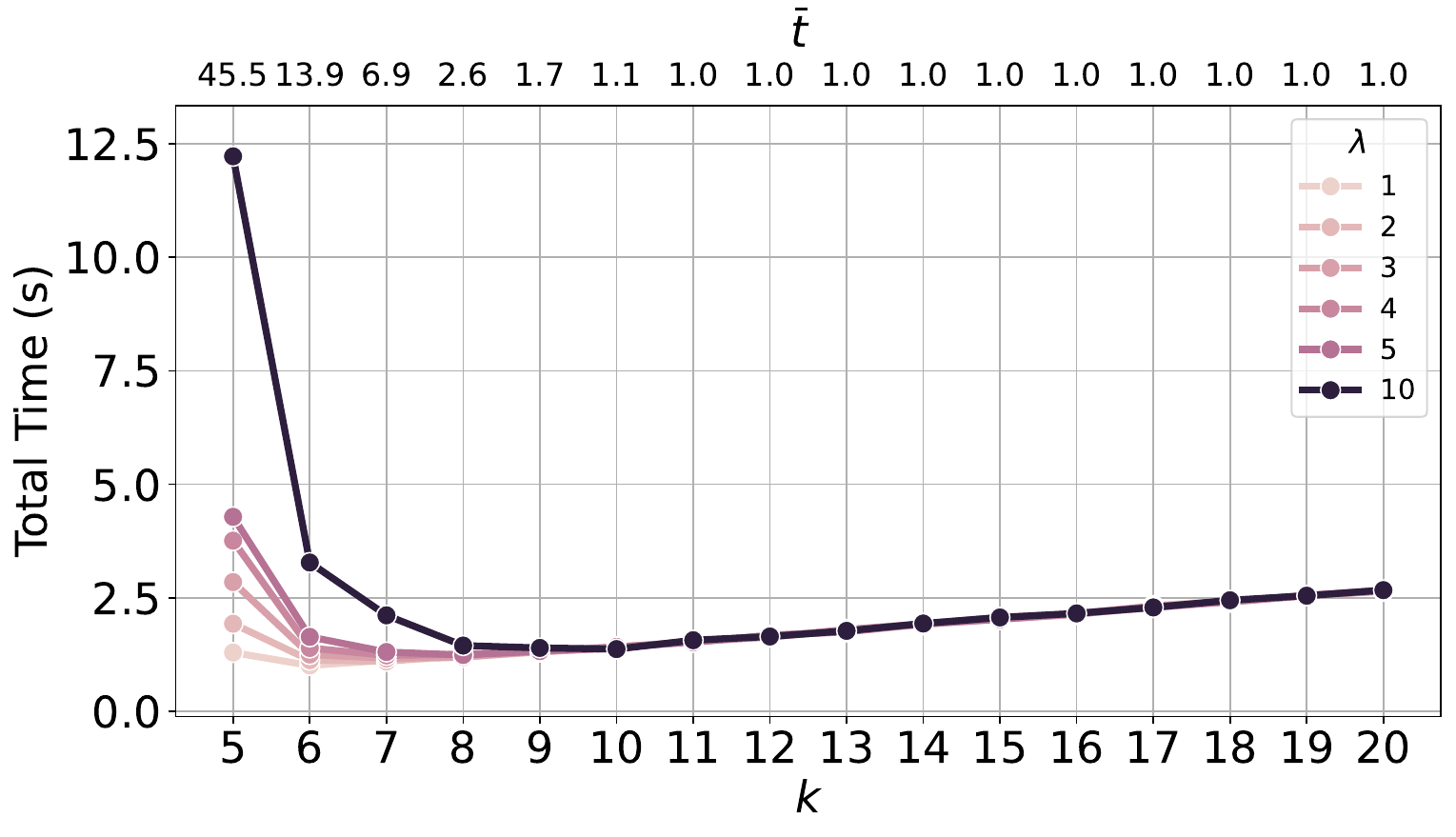}
         \caption{Shape}
         \label{fig:ShapelFitTime}
     \end{subfigure}
     \hfill
     \begin{subfigure}[b]{0.49\textwidth}
         \centering
         \includegraphics[width=\textwidth]{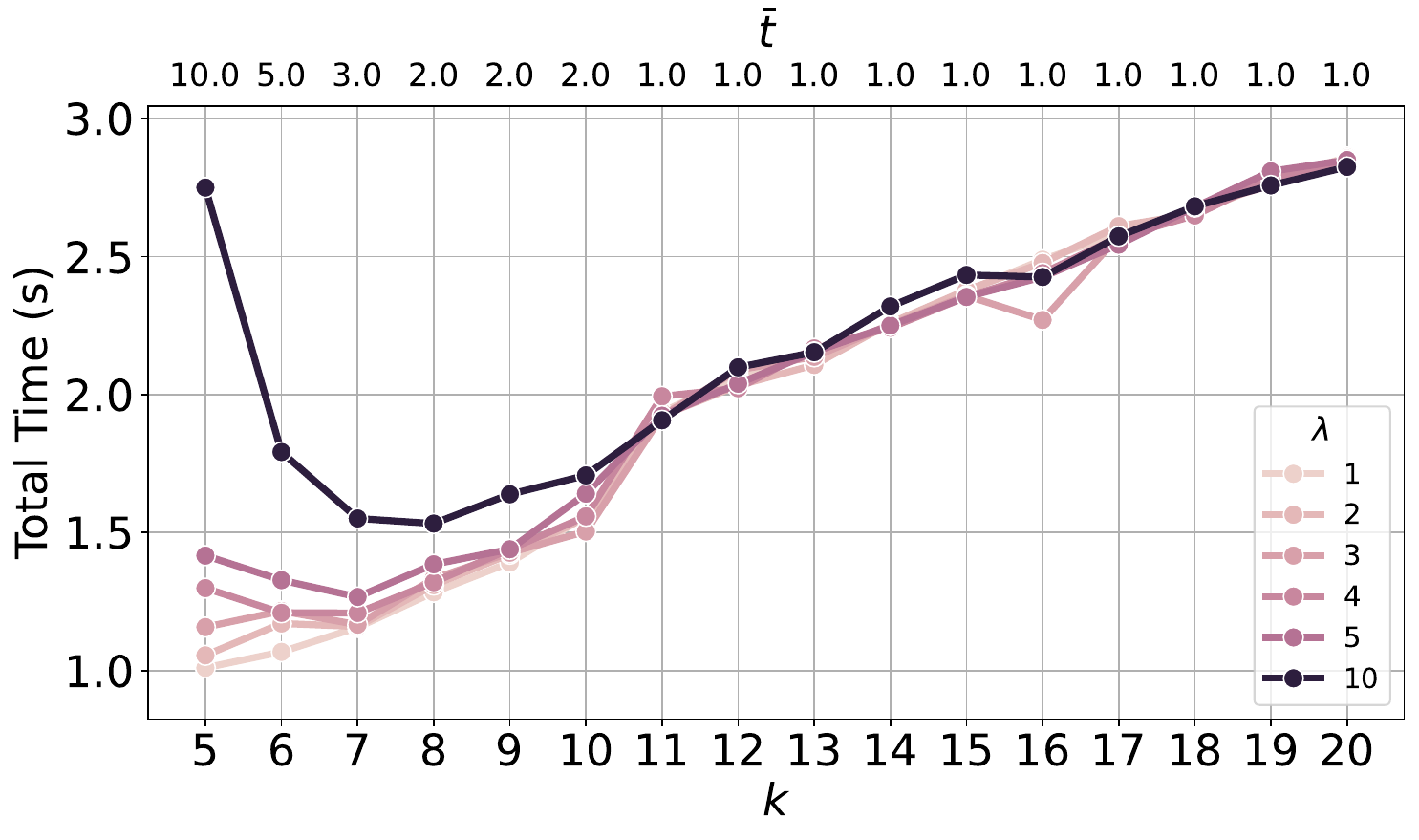}
         \caption{Shuttle}
         \label{fig:ShuttleFitTime}
     \end{subfigure}
     \hfill
     \begin{subfigure}[b]{0.49\textwidth}
         \centering
         \includegraphics[width=\textwidth]{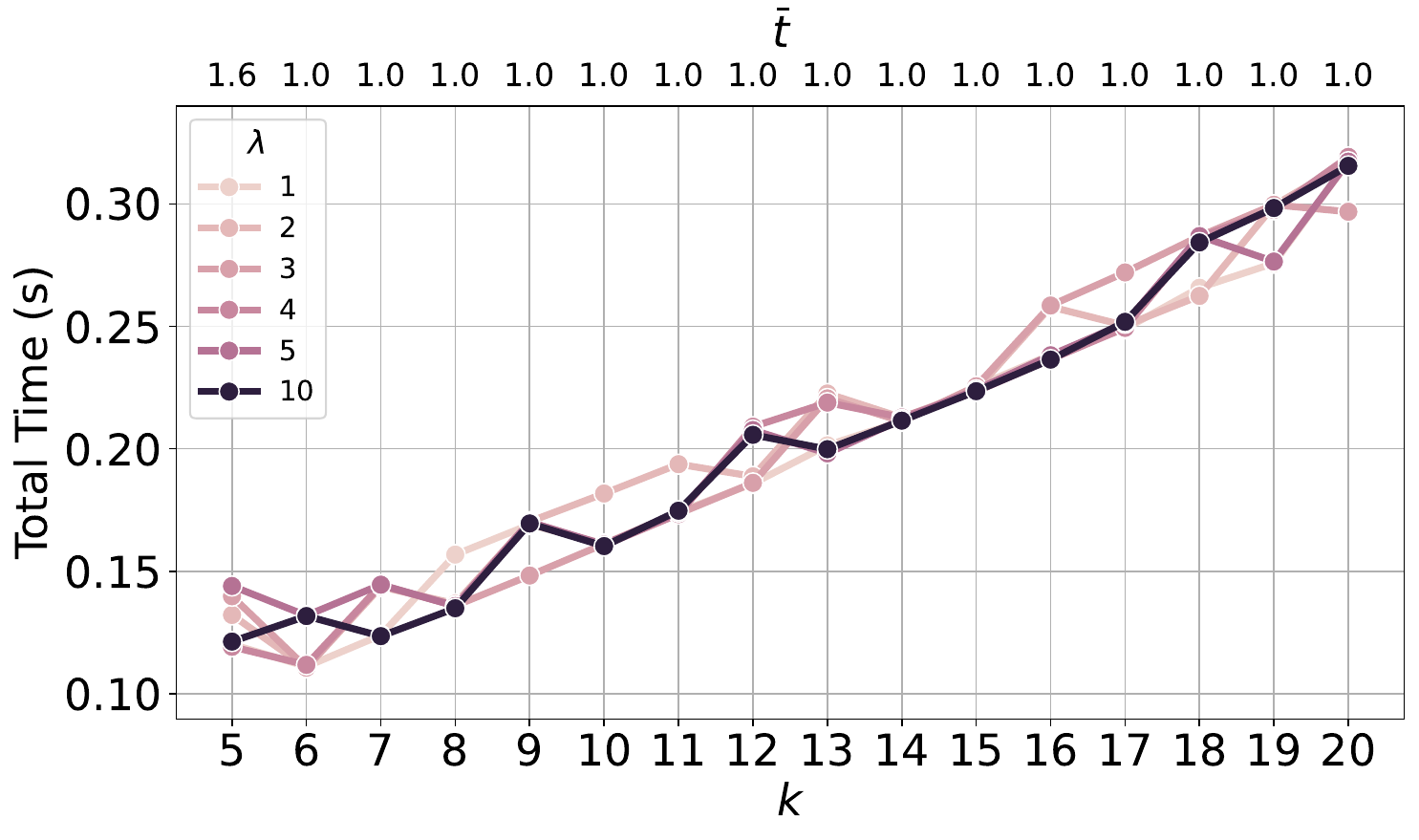}
         \caption{Speech}
     \end{subfigure}
     \hfill
     \begin{subfigure}[b]{0.49\textwidth}
         \centering
         \includegraphics[width=\textwidth]{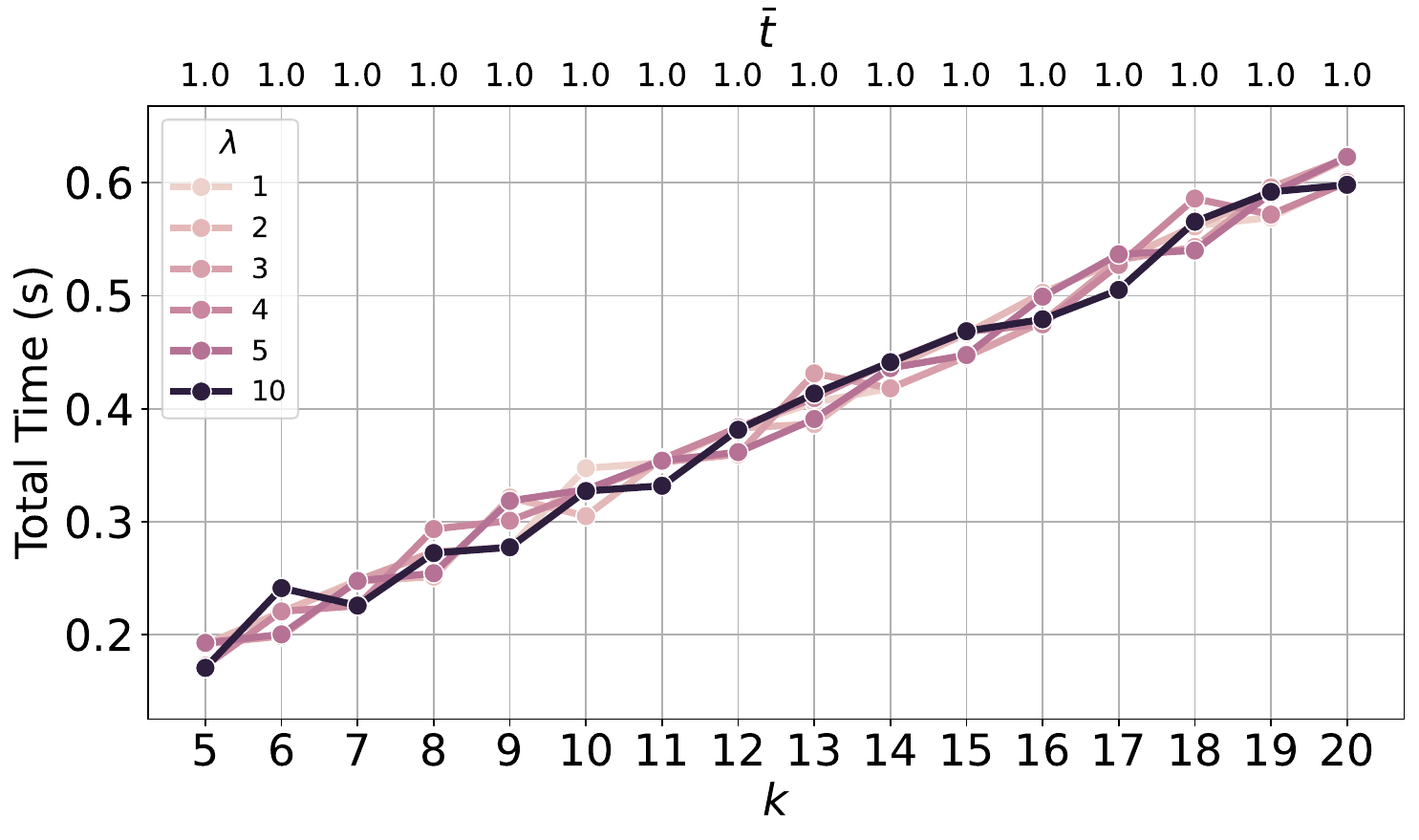}
         \caption{MNIST}
     \end{subfigure}
     \hfill
     \begin{subfigure}[b]{0.49\textwidth}
         \centering
         \includegraphics[width=\textwidth]{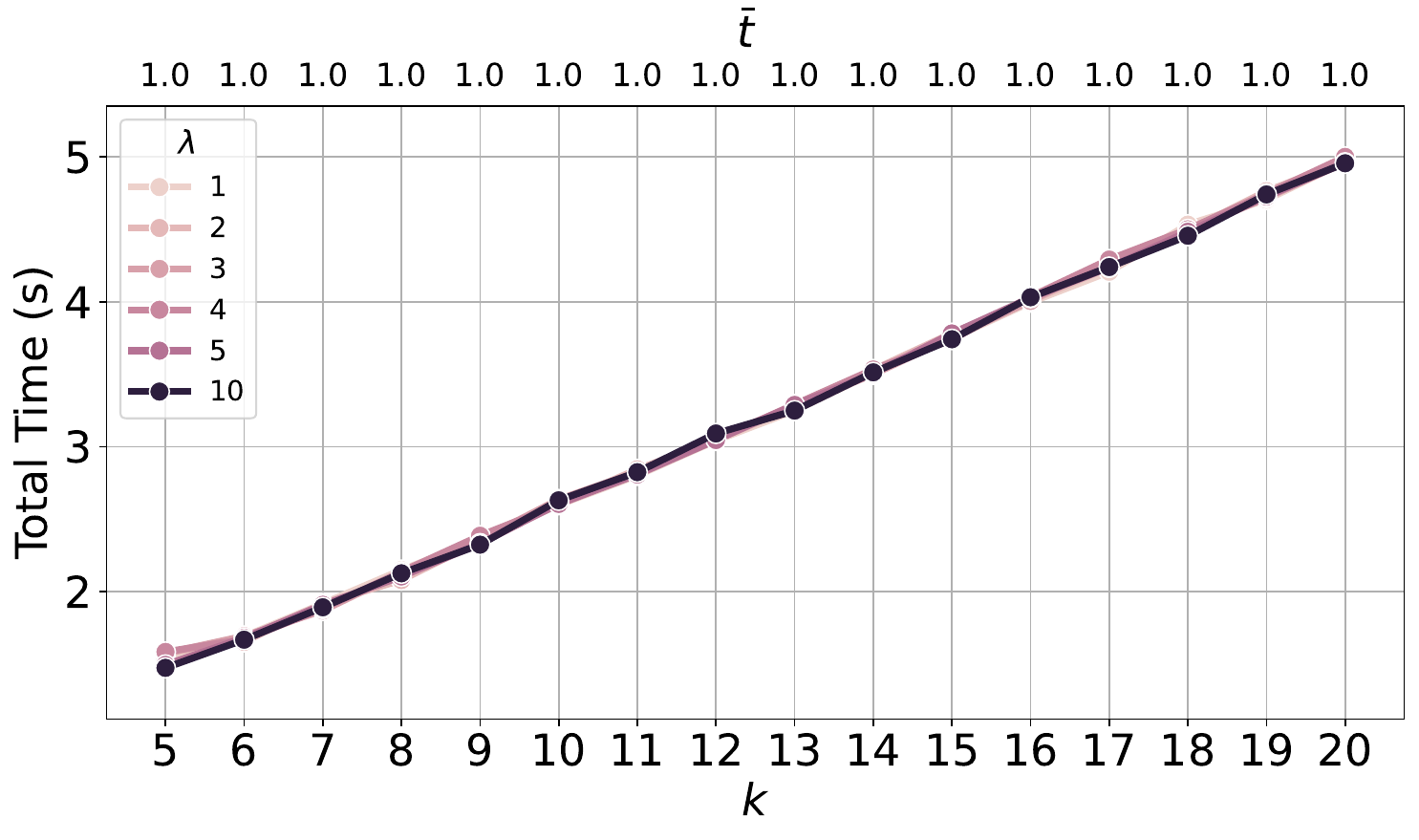}
         \caption{Audio}
     \end{subfigure}
     \hfill
     \begin{subfigure}[b]{0.49\textwidth}
         \centering
         \includegraphics[width=\textwidth]{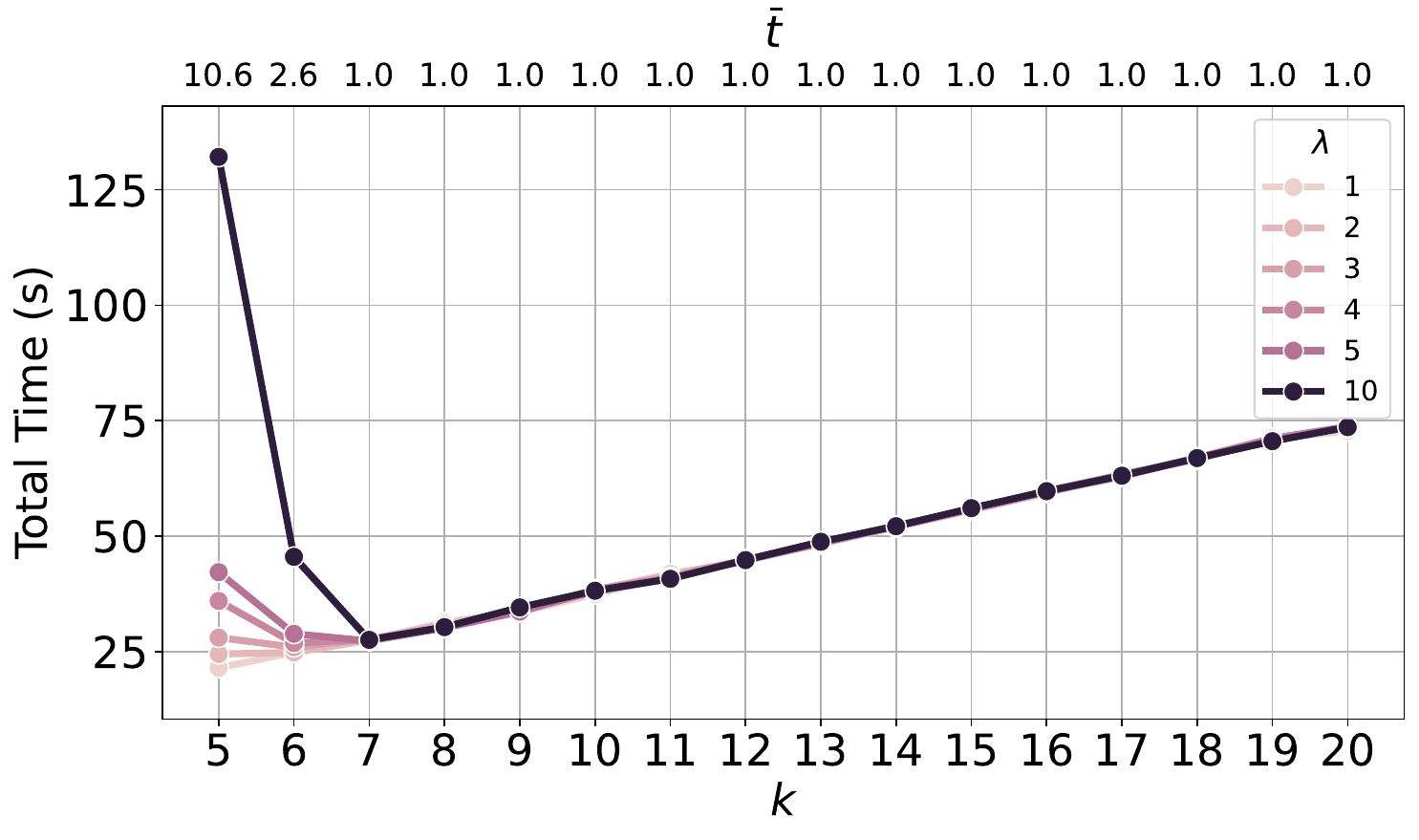}
         \caption{Corel}
         \label{fig:CorelFitTime}
     \end{subfigure}
     \hfill
     \begin{subfigure}[b]{0.49\textwidth}
         \centering
         \includegraphics[width=\textwidth]{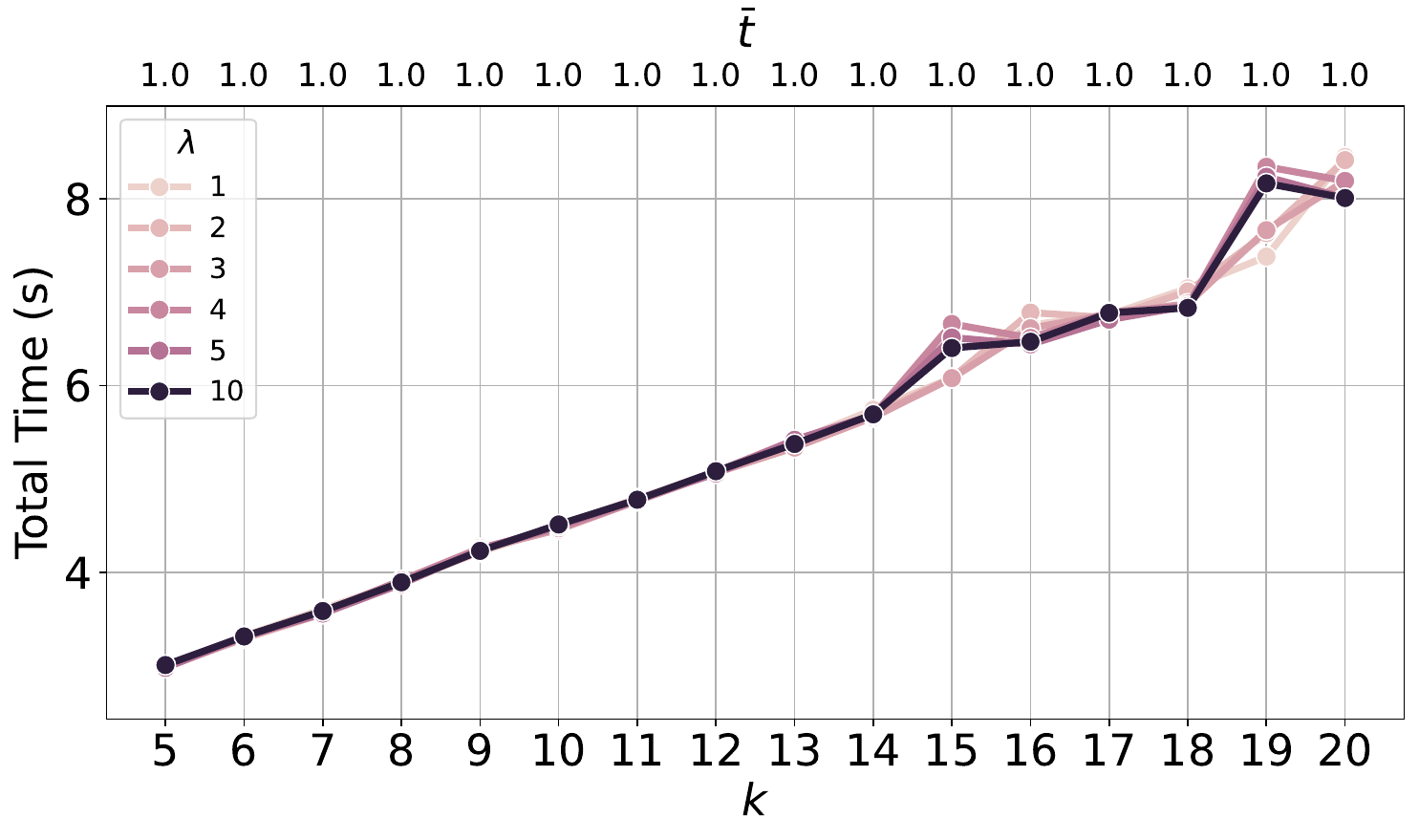}
         \caption{F-MNIST}
     \end{subfigure}
     \hfill
     \begin{subfigure}[b]{0.49\textwidth}
         \centering
         \includegraphics[width=\textwidth]{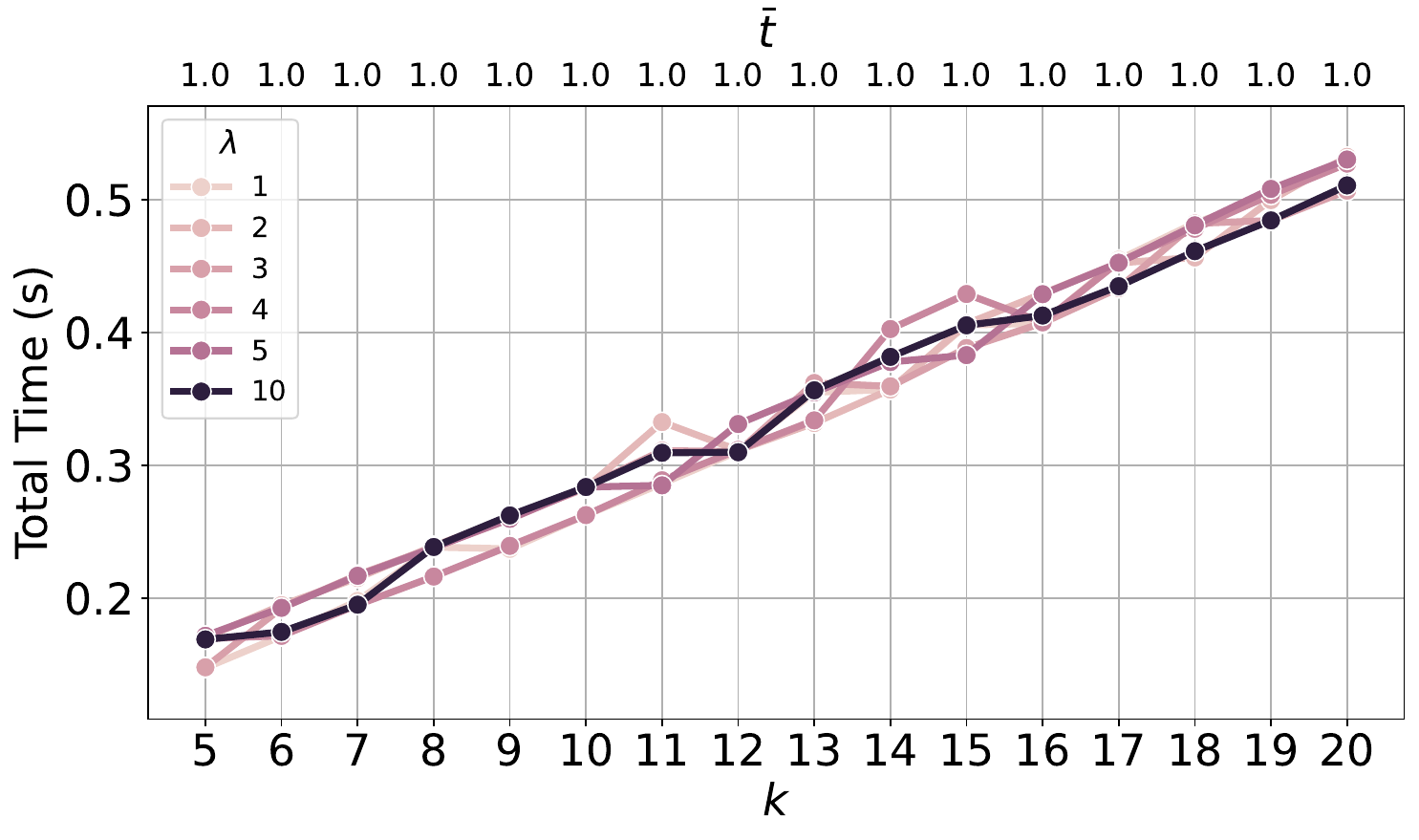}
         \caption{Miss America}
         \label{fig:missa1FitTime}
     \end{subfigure}
        \caption{Time complexity for different datasets using different $k$ and $\lambda$ values; $\bar{t}$ is the average of $t$ (number of disconnected components in the \ac{ANN} graph) over different runs }
        \label{fig:fitTimeLineGraph}
\end{figure*}

\paragraph{Effect on computation time}
Figure \ref{fig:fitTimeLineGraph} reveals the critical relationship between $k,~\lambda$, and computation time. The hyperparameter $k$ affects computation time in two ways: (1) it increases the cost of \ac{ANN} construction, but importantly, it also (2) reduces the number of disconnected components in the initial graph. When $k$ is small, the graph contains more disconnected components, requiring more inter-component edges.

This creates an important interdependency with $\lambda$. When $\lambda > k$ and $k$ are small values, computational time increases dramatically because the algorithm must process a larger number of inter-component edges. This effect is clearly visible in Figures \ref{fig:ShapelFitTime}, \ref{fig:ShuttleFitTime} and \ref{fig:CorelFitTime} (Corel dataset), where the computation time for $\lambda =10$ with $k =5$ is significantly higher than for $k = 10$ with the same $\lambda$ value.

The relationship between $k$ and computation time is approximately linear, consistent with our complexity analysis in Section \ref{sec:complexityAnalysis}.
As $k$ increases, the number of disconnected components naturally decreases, reducing the total number of inter-component edges that need processing during the refinement stage. This explains why the impact of $\lambda$ on computation time becomes less at higher $k$ values.

\paragraph{Optimal hyperparameter selection}
Based on these observations, we recommend the following hyperparameter selection strategy:

\begin{itemize}
    
    \item Keep $\lambda \le k$ to avoid excessive computational overhead. Setting $\lambda > k$, especially with small $k$ values, leads to inefficient computation due to a large number of inter-component edges.
    
    \item For optimal balance, we recommend $5 \leq k \leq 15$ with $2 \leq \lambda \leq 5$. This configuration provides excellent approximation quality while maintaining reasonable computation time.
    
    \item For very large datasets where computational efficiency is critical, setting $k =10$  and $\lambda = 5$ offers a good balance, as demonstrated in Table \ref{tab:datasets_and_results}.
     
\end{itemize}

\section{Further improvements}
\label{sec:furtherImprovements}
Section \ref{sec:Dimensionality} reveals that optimization efforts should primarily target the \ac{ANN} graph construction phase to achieve further performance improvements for large-scale datasets. Given that \ac{ANN} construction is the primary computational bottleneck, the integration of GPU-accelerated \ac{ANN} methods such as GNND \cite{wang2021fast} or GPU-HNSW \cite{zhao2020song} would likely yield substantial additional performance gains for \ac{FAMST}. 

Recent advancements in GPU-accelerated nearest neighbor search have demonstrated speedups of 50-180x over single-thread CPU implementations \cite{zhao2020song}, suggesting that such integration could potentially reduce the overall execution time of \ac{FAMST} by a similar order of magnitude. This presents a promising direction for future optimization, as our algorithm's modular design allows for seamless substitution of the \ac{ANN} construction component without affecting the subsequent refinement and \ac{MST} extraction phases.

The near-constant runtime of our refinement process across high-dimensional datasets of varying sizes highlights the efficiency of our local neighborhood exploration strategy and ensures it remains scalable, even for high-dimensional, large datasets.

\section{Conclusion}
\label{sec:conclusion}

This paper presents \ac{FAMST} that successfully addresses a fundamental computational bottleneck in data analysis by enabling efficient \ac{MST} construction for large-scale and high-dimensional datasets. With efficient scaling, consistent performance across diverse datasets, and minimal approximation error, the algorithm offers a practical solution for modern data analysis where exact \ac{MST} computation is infeasible. 

\section*{Acknowledgments}
This work was supported by the European Union within the framework of the Artificial Intelligence National Laboratory under Project RRF-2.3.1-21-2022-00004.


\begin{acronym}
\acro{FAMST}{Fast Approximate Minimum Spanning Tree}
\acro{$k$NN}{$k$-Nearest Neighbor}
\acro{ANN}{Approximate Nearest Neighbor}
\acro{MST}{Minimum Spanning Tree}
\acro{DFS}{Depth-First Search}
\acro{EMST}{Euclidean Minimum Spanning Tree}
\acro{NSW}{Navigable Small World}
\acro{HNSW}{Hierarchical Navigable Small World}
\acro{MFC}{Metric Forest Completion}
\end{acronym}

\bibliographystyle{ieeetr}
\bibliography{Reference}

\end{document}